\documentclass[12pt, letterpaper]{article}
\usepackage[utf8]{inputenc}
\usepackage[margin=1in]{geometry}

\usepackage[draft]{changes}
\newcommand{\stkout}[1]{\ifmmode\text{\sout{\ensuremath{#1}}}\else\sout{#1}\fi}
\setdeletedmarkup{\stkout{#1}}

\usepackage{amsmath, amsthm, amssymb, calrsfs, wasysym, verbatim, bbm, color, graphics, geometry, indentfirst, amsfonts}

\usepackage[american]{babel}
\usepackage{csquotes}
\usepackage[style=apa]{biblatex}
\usepackage{hyperref} 
\hypersetup{
    colorlinks=true,
    }
\usepackage{doi}

\newcommand{\dto}{\buildrel d \over \to}

\usepackage{array}
\usepackage{longtable}
\usepackage{booktabs}
\usepackage{tabularx}
\newcolumntype{E}{>{\raggedright\arraybackslash}X}
\newcolumntype{I}{>{\raggedleft\arraybackslash}X}
\usepackage{tabulary}

\usepackage{chngcntr}
\usepackage{titling}

\usepackage{graphicx}
\usepackage{subcaption}
\DeclareCaptionLabelSeparator{mytable}{.\\}
\usepackage{enumerate}

\theoremstyle{definition} 

\newtheorem*{remark*}{Remark} 
\newtheorem{assumption}{Assumption}
\newtheorem{theorem}{Theorem}

\newcommand{\Y}{\mathbf{Y}}

\newcommand{\D}{\mathbf{D}}

\newcommand{\iid}{\stackrel{iid}{\sim}}

\newcommand{\ind}{\mathbf{1}}

\DeclareMathOperator*{\argmin}{arg\,min}

\usepackage{mathtools}

\usepackage{setspace} 

\usepackage{authblk}

\title{A Partial Linear Estimator for Small Study Regression Discontinuity Designs}
\author[1]{Daryl Swartzentruber} 
\author[2]{Eloise Kaizar}
\affil[1]{Departments of Data Science and Mathematics, Centre College}
\affil[2]{Department of Statistics, The Ohio State University}
\date{March 6, 2025}

\begin{document}

\maketitle

\begin{abstract}
Regression discontinuity (RD) designs are a popular approach to estimating a treatment effect of cutoff-based interventions. Two current estimation approaches dominate the literature. One fits separate regressions on either side of the cutoff, and the other performs finite sample inference based on a local randomization assumption. Recent developments of these approaches have often focused on asymptotic properties and large sample sizes. Educational applications often contain relatively small samples or sparsity near the cutoff, making estimation more difficult. As an alternative to the aforementioned approaches, we develop a partial linear estimator for RD designs. We show in simulations that our estimator outperforms certain leading estimators in several realistic, small-sample scenarios. We apply our estimator to school accountability scores in Indiana.
\end{abstract}

\section{Introduction}

Regression discontinuity (RD) designs are a quasi-experimental framework originally developed by \textcite{Thistlewaite}. In the classic RD design a treatment is assigned to all individuals who score above a certain cutoff of a continuous running variable, and linear regression is used to estimate the causal effect of the treatment. The prevalence of cutoff-based interventions in the fields of economics and education have made RD designs a popular causal inference method in recent decades and have led to important methodological advances. 

Currently two approaches dominate the RD literature. The first, popularized by \textcite{hahn2001}, relies on assumptions about the continuity of the underlying mean function. Practitioners typically fit separate models on either side of the cutoff, and estimate the local average treatment effect (LATE) as the difference between the values of the response predicted at the cutoff based on the two models. Nonparametric regression is often preferred due to the local nature of RD designs \parencite{Gelman2019}, necessitating that users choose a tuning parameter called a bandwidth. Popular algorithms calculate bandwidths optimized for different inferential methods, including those proposed by \textcite{IK2011} (henceforth IK), \textcite{CCT2014} (henceforth CCT), and \textcite{armstrong2020simple} (henceforth AK). 

The second approach to RD estimation was popularized by \textcite{cattaneo2015randomization}, who rely on local randomization, i.e., the idea that observations near the cutoff fall on either side of the cutoff essentially at random. Practitioners typically choose a window near the cutoff where the local randomization is assumed to hold and then use finite sample methods to estimate the LATE. Much like  for bandwidths in the continuity approach, much of the current research in this area focuses on algorithms to automatically choose the window.

Much recent methodological development has focused on economic applications, where the sample sizes are typically quite large. \textcite{swartzentruber2024}, henceforth SK, note that educational applications often feature smaller sample sizes or samples that are sparse around the cutoff. SK compare the performance of popular RD estimation methods within both major approaches. Simulation results show that some methods known to have desirable asymptotic qualities do not work particularly well with small or sparse samples.

In this paper, we propose a method that outperforms currently popular RD estimation methods in many small sample scenarios. Our method is an implementation of the partial linear RD estimator explored by \textcite{porter2003estimation}, henceforth Porter. The partial linear approach has been neglected in the literature, perhaps because Porter shows some asymptotic inferiorities compared to the approach that fits separate models on either side of the cutoff. We argue that these asymptotic results are not relevant for small or sparse samples, for which the partial linear approach has certain desirable qualities. We develop a tailored bandwidth selection algorithm and variance estimation technique and use simulation to show that our partial linear estimator tends to have better operating characteristics than competing methods in several realistic small sample scenarios. 

The paper is organized as follows. Section \ref{ple} describes the local polynomial estimation method that dominates the continuity approach, and contrasts it with Porter's original partial linear estimator. Section \ref{implementation} highlights the modifications we make to Porter's estimator, including the details of our bandwidth selection algorithm and variance estimation technique. In Section \ref{simstudy} we present the results of a simulation study comparing our method to several competing methods. We provide a real-data example using school accountability scores in Section \ref{application} before concluding in Section \ref{conclusion}.

\section{Overview of Select RD Estimation Methods} \label{ple}

In this paper, we focus on the sharp RD design, in which treatment $D$ is based solely on the relationship between the value of the continuous running variable $X$ and the cutoff $c$, namely $D=\ind_{[X\geq c]}$. Let $Y$ be a continuous response variable. We define the true underlying mean function as $E(Y|X)=\mu(x)=\mu^-(x)\ind_{[X<c]}+\mu^+(x)\ind_{[X\geq c]}$, where $\mu^-(x)$ and $\mu^+(x)$ are the mean functions below and above the cutoff, respectively. We are interested in estimating the LATE, $\tau=\mu^+(c)-\mu^-(c)$. 

\textcite{hahn2001} advocate for the use of local polynomial regression to separately estimate $\mu^-(x)$ and $\mu^+(x)$ and use these to estimate $\tau$. This nonparametric method uses a kernel function and a bandwidth to estimate the mean at a given point as a weighted average of the response values close to that point. Here we will refer to this method as local polynomial estimation (LPE). Local constant regression (LPE with polynomial degree 0) is known to suffer from bias when estimating near boundary points, and thus \textcite{hahn2001} advocate the use of local linear regression as a way to mitigate this bias.  

As an alternative to LPE, Porter considers an approach based on partial linear models, which are semiparametric models with both a parametric and a nonparametric component. Porter recognizes that an RD model can be written as
\begin{equation} \label{eq:partiallinear}
    Y_i=\tau D_i +\mu^*(X_i)+ \epsilon_i,
\end{equation}
where $\mu^*(X_i) = \mu(X_i)-\tau D_i$ is the underlying nonparametric mean function with the discontinuity $\tau$ subtracted from points above the cutoff. Porter rewrites the model as $Y_i-\tau D_i=\mu^*(X_i)+\epsilon_i$, allowing $Y-\tau D$ to become the response variable. He can then write out an expression for the least squares criteria and estimate $\tau$ by minimizing that objective function:
\begin{equation} \label{tauhatporter}
    \hat{\tau}=\argmin_{\tau}\sum_{i=1}^n\left[y_i-\tau d_i -\sum_{j=1}^n l_j(x_i)(y_j-\tau d_j)\right]^2.
\end{equation}
where $l_j$ are the weights used in the nonparametric regression calculated via a kernel function and bandwidth.

This estimator is equivalent to one proposed for general partial linear models by \textcite{robinson1988}, who shows that Equation \ref{eq:partiallinear} implies that $Y_i-E(Y|X_i)=\tau(D_i-E(D|X_i)) + \epsilon_i$. He uses the residuals from two constant regression estimations, $Y_i-\hat{E}(Y|X_i)$ and $D_i-\hat{E}(D|X_i)$ to estimate $\tau$ by applying a no-intercept OLS model and proved the consistency of the resulting estimator. Porter shows that different consistency properties arise if $D$ is a deterministic function of $X$, as for the RD model, but the conceptualization of \textcite{robinson1988} motivates the variance estimation technique we develop in Section \ref{implementation}.

One important distinction between this semiparametric estimator and the currently popular two-model estimator is that, for PLE, the behavior of $\mu^*(X)$ at the cutoff $x=c$ should follow any constraints, such as ``smoothness", imposed by the mean function estimator chosen by the practitioner. Thus, PLE estimators use data from both sides of the cutoff to estimate the underlying mean function. One potential advantage of using a single nonparametric mean function is that this estimator is not vulnerable to the same boundary bias as the estimator that uses two separate local polynomial regression models. Exploring this feature of PLE, Porter derives asymptotic distributions of $\hat{\tau}$ based on local constant regression under both a stronger and weaker set of smoothness assumptions for the underlying mean function and uses asymptotic variance to develop plug-in standard error formulae. 

Porter compares the asymptotic properties of the LPE and his partially linear estimator (PLE). He shows that when the stronger smoothness assumption holds, both the PLE and the LPE achieve optimal rates of convergence, but the PLE does not in general achieve the optimal rate when only the weaker smoothness assumption holds. The LPE is able to achieve the optimal rate under the weaker assumption, although the degree of the local polynomial needed depends on the amount of smoothness. He concludes by advocating for the use of the LPE based on potential robustness, a point reiterated by \textcite{vanderklaauw2008}. Perhaps due to this recommendation, the PLE has been given scant attention in the literature over the last 20 years.

\section{Implementation of the Partial Linear Estimator} \label{implementation}

While the partial linear model may not be asymptotically superior, the PLE's reliance on smoothness at the cutoff and use of data on both sides of the cutoff  may mitigate the limited statistical power due to sparse data near the cutoff. However, unlike for LPE where a rich array of nonparametric models have been explored, to our knowledge the local constant regression is the only nonparametric model that has been fully examined and implemented for use with PLE for RD analyses. In this paper, we fill this gap in the literature by proposing broader options for the weights, along with a customized bandwidth algorithm and method for approximating the standard error.

Porter shows that using weights based on local constant regression with PLE does not suffer from the poor boundary properties that would come from using such weights in the two-model approach. While Porter suggested the use of higher order kernels for further bias reduction in certain situations, he did not pursue this avenue to full implementation. \textcite{marron1994higherorder} notes that asymptotic gains from the use of higher order kernels in nonparametric regression often require quite large sample sizes, and that there are other drawbacks such as the poor interpretability of negative weights. For these reasons, and to be more in line with the rest of the RD estimation literature, we also do not pursue higher order kernels here. Rather, motivated by the relationship between higher order kernels and higher order local polynomial regression (see \textcite{Yu2016} and \textcite{fan1997local}), we modify Porter's estimator in Equation \ref{tauhatporter} to use local polynomial weights $l = l^{LP}$ of a general degree $p$, and denote the resulting effect size estimator $\hat{\tau}^{PLE}$. 

The solution to the minimization is easily seen in matrix form. Let $\Y$ be the $n \times 1$ vector of responses. Let $l^{(i)}=(l_1^{LP}(x_i),...,l_n^{LP}(x_i))'$ be the vector of weights for $x_i$, and let these vectors be combined into the $n \times n$ weight matrix $\mathbf{L}=(l^{(1)},...,l^{(n)})$. Let $\mathbf{D}=(\ind_{\{x_1 \geq c\}},..., \ind_{\{x_n \geq c\}})$ be a $n\times 1$ vector of treatment indicators. Finally, let $\mathbf{G}=(\mathbf{I}-\mathbf{L^T})\mathbf{D}$. Then, it can be shown that 
\begin{equation} \label{tauhatple}
  \hat{\tau}^{PLE}=\mathbf{(G'G)}^{-1}\mathbf{G'(I-L')Y},  
\end{equation}
which is a linear combination of the observed responses. 

In addition to the polynomial degree, the weights depend on the choice of kernel and bandwidth. Following the advice from \textcite{fan1996local}, LPE methods often employ the triangular kernel because it is thought to be optimal for boundary estimation. But since the cutoff is no longer at a boundary in the PLE method, we focus on the use of the optimal interior estimation Epanechnikov kernel to calculate the local polynomial weights.

\subsection{PLE Bandwidth Selection Algorithm} \label{skadevelop}
Porter does not provide a bandwidth selection algorithm optimized for PLE with local constant weights. He suggests that cross-validation could be used, but this is no longer a common RD bandwidth selection approach, and it would be difficult to apply in a small sample setting. One could certainly pair other popular bandwidth selection algorithms with PLE. However the popular bandwidth algorithms from IK and CCT rely on the assumption that the second derivatives of $\mu^*(X)$ differ at the cutoff.  But recall that the motivation for the PLE involves estimating a single smooth mean function $\mu^*(X)$, from which the discontinuity is subtracted from points receiving the treatment. Thus, we make use of the stronger of Porter's two smoothness assumptions:

\begin{assumption} \label{sma}
    Right and left-hand derivatives of $\mu^*(X)$ to order $\lambda_{\mu}$ are equal at c.
\end{assumption}

We expect that in most educational applications we will not see substantial deviations from this smoothness assumption, and we believe that the PLE approach would still perform well in small sample regimes with modest deviations from Assumption \ref{sma}.

Porter proves the following result about $\hat{\tau}$ with local constant functions and second-order kernels, which is equivalent to $\hat{\tau}^{PLE}$ with $p=0$.
\begin{theorem} \label{asymp dist}
    If Assumption \ref{sma} holds with $\lambda_\mu \geq 3$ and $h^3\sqrt{nh}\to A$, where $0\leq A <\infty$, then
    \begin{equation*}
        \sqrt{nh}(\hat{\tau}-\tau) \dto N\left(Ab_P,C_{P1}\frac{\sigma^2_+(c)+\sigma^2_-(c)}{4f(c)}\right),
    \end{equation*}
    where $b_P=2K_2(0)\left(f(c)\int_0^\infty K_0^2(w)dw\right)^{-1}\left(\frac{f'(c)}{f(c)}g_2(c)\int_0^\infty K_1(v)dv -g_2'(c)\int_0^\infty K_0(v)vdv \right)$,\\
     $g_2(x)=\mu^{*'}(x)f'(x)+\mu^{*''}(x)f(x)/2$, \\
    and $K_0$, $K_1$, $K_2$, and $C_{P1}$ are functions of the chosen kernel.
\end{theorem}
From this distribution, we derive the asymptotic mean squared error (AMSE) of $\hat{\tau}$ with local constant regression to be 
\begin{equation} \label{mseple}
    AMSE_{PLE}(h)=h^6 b_P^2 + \frac{C_{P1}(\sigma^{2}_+(c)+\sigma^{2}_-(c))}{4nhf(c)}.
\end{equation}
By minimizing this expression with respect to the bandwidth $h$ we arrive at an infeasible optimal bandwidth, which we denote $h_{SM}$ in reference to the smoothness assumption on which it relies:
\begin{equation} \label{hska}
    h_{SM}=\argmin_h  AMSE_{PLE}(h)=\left(\frac{C_{P1}(\sigma^{2}_+(c)+\sigma^{2}_-(c))}{24nb_P^2f(c)}\right)^{1/7}.
\end{equation}
For a fully data-driven plug-in bandwidth $\hat{h}_{SM}$, we must estimate the unknown quantities in the expression above. For this estimation we often modify ideas from the plug-in approaches of IK, CCT, and \textcite{AI2018} (henceforth AI), which require the estimation of similar quantities. The plug-in estimands fall into three main categories: the density of the running variable and its derivatives, the derivatives of the underlying mean function, and the variance function. We address each category in turn.

\subsubsection{Estimating the Density of the Running Variable and its Derivatives} \label{est den}
The expression for $b_P$ involves the density of the running variable and its first and second derivatives evaluated at the cutoff. As discussed in \textcite{JonesKDDE}, kernel density estimation and kernel density derivative estimation are fairly standard ways to estimate the density of an unknown function and its derivatives, respectively. 

For some bandwidth $h$ and kernel $K$, we can estimate the density $f(x)$ with
\begin{equation} \label{kde}
    \hat{f}(x)=\frac{1}{n}\sum_{i=1}^n\frac{1}{h}K\left(\frac{x-X_i}{h}\right).
\end{equation}
To calculate derivatives of the density at a point, we choose to follow the seemingly most popular and widely implemented approach, which is to simply differentiate the right hand side of Equation \ref{kde}. This approach requires the kernel to be piece-wise differentiable on its support and the derivative to be non-constant on that support. To meet these requirments, we employ the Gaussian kernel, $K(x)=\frac{1}{\sqrt{2\pi}}e^{-\frac{1}{2}x^2}$, rather than the Epanechnikov or triangular kernels often used in RD estimation. For consistency we use this kernel for the density estimation as well. We implement this estimation using the \textit{kdde} function in the \textit{ks} package \parencite{ks} in R. We use the well-known plug-in estimator of \textcite{wandjones} to choose the bandwidth.

\subsubsection{Estimating Derivatives of the Mean Function} \label{est deriv}
To estimate $b_P$ we also need estimates of the first three derivatives of the underlying mean function. This is perhaps the most challenging of the quantities needed, especially considering that our ultimate goal involves estimating the underlying mean function itself. IK, CCT, and AI propose slightly different multi-stage approaches that make use of local polynomial techniques. We incorporate some of their ideas in our estimation, but one important distinction is that since we are assuming the derivatives of the mean function are the same on both sides of the cutoff, we need one overall estimate for each derivative rather than one on each side. 

We take as our starting point the infeasible optimal bandwidth formula for estimating the $\nu$th derivative of a function at a point in the interior of the function's support from \textcite{fan1996local},
\begin{equation} \label{fg band}
    h_{FG}=C_{FG}(\nu,\rho,K)\left(\frac{\sigma^2(c)}{(\mu^{*^{(\rho+1)}})^2f(c)n}\right)^{1/(2\rho +3)},
\end{equation}
where $\rho$ is the degree of the local polynomial we are using to approximate the underlying mean function, and $C_{FG}(\nu,\rho,K)$ is a constant that depends on $\nu$, $\rho$, and the kernel $K$.  This formula is optimal when $\rho -\nu$ is odd. We choose to use $\rho=\nu+1$ for $\nu \in \{1,2,3\}$. 

For each value of $\rho$, we need an estimate of the derivative $\mu^{*^{(\rho+1)}}$. We achieve this in a similar manner to IK, by fitting global polynomials that include a treatment indicator. \textcite{fan1996local} suggest fitting a polynomial of degree $\rho+3$ but we feel that such a high degree would not be ideal in a small sample scenario. Instead we follow the strategy of CCT and fit polynomials of degree $\rho+1$. Thus we fit global polynomials with a jump at the cutoff of degree 3, 4, and 5 (corresponding to $\rho=$2, 3, and 4, respectively), and from these polynomials we calculate $\mu^*{^{(\rho+1)}}$ for each $\rho$. 

Equation \ref{fg band} also requires a plug-in estimate of the variance function at the cutoff, $\sigma^2(c)$. We mirror the approach of AI to use the estimated variance from the polynomial regression, which assumes homoscedasticity, rather than the estimate of the variance function in Section \ref{est var}. However, for $f(c)$ we do use the same value estimated in Section \ref{est den}. Plugging in these quantities gives data-driven pilot bandwidths $\hat{h}_{FG}$ for each of $\nu \in \{1,2,3\}$. 

IK, CCT, and AI all use their pilot bandwidths in subsequent local polynomial regressions on either side of the cutoff as described above. However, we are unable to use traditional local polynomial estimation at this stage because we are estimating a single mean function with a structurally imposed discontinuity. Rather, we incorporate ideas from both parametric and nonparametric methods. We restrict our data to just those observations within one bandwidth of the cutoff, essentially using a uniform kernel similar to the approach of IK. However, rather than fitting a regular polynomial like IK does, we fit a polynomial with a treatment indicator on this subset of our data similar to the polynomial we fit globally in the first stage of our approach. We choose the degree of the polynomial we fit at this stage to be equal to the value of $\rho$. From these polynomials we calculate the desired derivatives $\mu^{*^{(\nu)}}(c)$ for $\nu \in \{1,2,3\}$.

\subsubsection{Estimating the Variance Function} \label{est var}
The numerator of $h_{SM}$ involves $\sigma^{2}_+(c)$ and $\sigma^{2}_-(c)$, the right and left-hand limits of the variance function at the cutoff. Note that we are not assuming here a constant variance function and thus desire separate calculations using data on either side of the cutoff. AI use variance estimates obtained from their estimation of the mean function derivatives described in Section \ref{est deriv}. However, we assumed homoscedasticity that stage and do not want to plug in that single variance estimate for both values. 

We use a nearest-neighbor approach similar to CCT in the hope that it will lead to more stable estimates for small samples. Thus we estimate the variance below and above the cutoff as 
\begin{equation} \label{nn var below}
    \hat{\sigma}^{2}_-(c)=\frac{1}{J-1}\sum_{i=1}^J (Y_{i,-}-\bar{Y}_{-})^2
\end{equation}
and
\begin{equation} \label{nn var above}
    \hat{\sigma}^{2}_+(c)=\frac{1}{J-1}\sum_{i=1}^J (Y_{i,+}-\bar{Y}_{+})^2,
\end{equation}
where $J$ is the number of nearest neighbors, $Y_{i,-}$ and $Y_{i,+}$ are the response values for the $J$ unique observations with $X_i$ values closest to the cutoff, below and above the cutoff, respectively, and $\bar{Y}_{-}$ and $\bar{Y}_{+}$ are the average of the J response values below and above the cutoff, respectively. For our implementation we follow CCT to choose J=3. 

\subsection{PLE Variance Estimation} \label{plevariance}

To conduct inference in the RD setting using our PLE estimator, we must choose a variance estimation technique. Porter proposes variance estimation based on the asymptotic distribution of the PLE estimator. However, for small studies this asymptotic distribution may be quite different from the true distribution of $\hat{\tau}$. We advocate for using an approach based on jackknife resampling. In Section \ref{appvar} of the supplemental appendix we present alternative approaches based on jackknife resampling as well as a direct plug-in approach, and show via simulation that our favored jackknife approach performs relatively well.

The jackknife is a resampling procedure that typically removes one observation at a time from a data set, calculates an estimate based on the remaining $n-1$ observations, and then utilizes those $n$ point estimates to estimate the variance of the overall estimate. \textcite{you2003jackknife} estimate the variance by applying a jackknife formula from \textcite{hinkley1977jackknifing} designed for a linear model to the parametric second step of the general partial linear estimation technique of \textcite{robinson1988}. This approach involves deleting one pair of residuals $\left(Y_i-\hat{E}(Y|X_i), D_i-\hat{E}(D|X_i)\right)$ each time, rather than one original observation, in order to estimate the parametric coefficient $\tau$.  We modify their method, using a jackknife formula from \textcite{wu1986jackknife} that has been shown to have certain advantages over Hinkley's formula. The result is our PLE variance estimator, 

\begin{equation} \label {wu r}
    \hat{V}_{PLE}(\hat{\tau}^{PLE})=\sum_{i=1}^n(1-w_i)(\hat{\tau}^{PLE}_{(i)}-\hat{\tau}^{PLE})^2=(\hat{\D}'\hat{\D})^{-2}\sum_{i=1}^n\frac{r_i^2}{1-w_i}\hat{d}_i^2,
\end{equation}
where $\hat{d}_i=d_i-\hat{E}(d|x_i)$, $\hat{\D}=(\hat{d}_1,...\hat{d}_n)'$, $r_i=y_i-\hat{E}(y|x_i)-\hat{d}_i\hat{\tau}$, $w_i=\hat{d}_i'(\hat{\D}'\hat{\D})^{-1}\hat{d}_i$, and $\hat{\tau}^{PLE}_{(i)}$ is the treatment effect estimate with the $i$th pair of residuals removed. 

This choice of variance estimation allows us to approximate a PLE $\left(1-\alpha\right)100\%$ confidence interval as
\begin{equation}
    I_{PLE}=\hat{\tau}^{PLE}\pm z_{\alpha/2}SE^{PLE}(\hat{\tau}),
\label{PLE_I}
\end{equation}
where $SE^{PLE}(\hat{\tau})=\sqrt{\hat{V}_{PLE}}$.

\section{Simulation Study} \label{simstudy}
In this section we present the results from a Monte Carlo simulation study comparing the operating characteristics of PLE methods developed in Section \ref{implementation} to several leading RD estimation methods that SK identified through simulation as performing well with small or sparse samples. 

We first consider the conventional continuity method in which a point estimate is calculated by fitting separate local linear regressions with triangular kernel on either side of the cutoff where the bandwidth is selected via the MSE-optimal bandwidth algorithm designed for it by IK. We refer to this method as CV/IK. 

Our second conventional comparator method is based on the idea of fixed-length confidence intervals (FLCI) of AK. They use the same point estimation approach as above, but inflate the critical value in order to achieve better confidence interval coverage. They also develop a bandwidth algorithm that pairs with their inferential technique. Their formulas rely on the second derivative bound of the underlying mean function, but they also propose a data-driven approximation of this value, which we use in our simulation. We refer to this method as FLCI/AK. 

We also consider the local randomization methods of \textcite{cattaneo2015randomization}, which rely on finite sample inference in a window close to the cutoff. They propose a window selection method that is not feasible for the simulation that we are performing here. Thus we follow the approach of SK and choose a window based on guaranteeing a minimum number of observations within that window. We use both five and ten for this minimum, and refer to these methods as LR/LR5 and LR/LR10, respectively. 

In addition to these existing methods we consider methods based on the PLE inferential approach given by equations \ref{tauhatple} and \ref{PLE_I}, using $\rho=1$ which corresponds to local linear estimation. Here we report the results of this approach paired with two bandwidth algorithms. The first is our SM bandwidth developed in Section \ref{skadevelop}. The second is the IK bandwidth, which we will show works well with the PLE approach despite the conflicting assumptions mentioned earlier. We refer to these methods as PLE/SM and PLE/IK, respectively. In both cases we use the Epanechnikov kernel.

We use R (v4.4.0; \textcite{R}) for all calculations. We use the package \textit{rdrobust} (v2.2; \textcite{rdrobust}) for the CV/IK method, the package \textit{RDHonest} (v1.0.0; \textcite{RDHonest} and v0.3.2; \textcite{RDHonestold}) for the FLCI/AK method, the package \textit{rdlocrand} (v1.0; \textcite{rdlocrand}) for the LR methods, and the package \textit{rdple} (v1.0.0; \textcite{rdple}) for the PLE methods.

The supplemental appendix shows the results from several other methods that did not perform as well. 

\subsection{Simulation Study Settings}

\begin{figure}[t!]
    \centering
    \includegraphics[width=\textwidth]{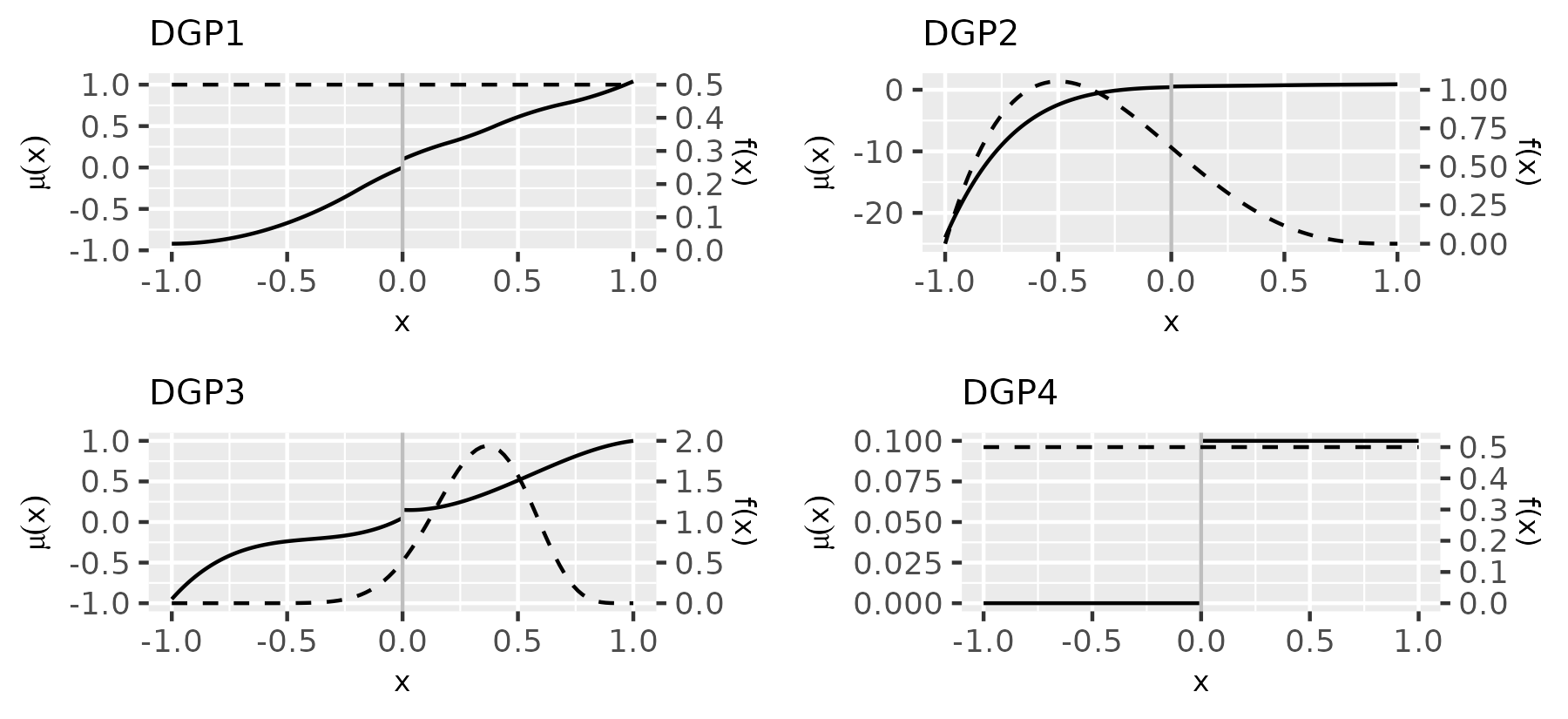}
    \caption{Mean function $\mu(x)$ (solid line) and density $f(x)$ (dashed line) of the DGPs.}
    \label{dgp.paper2}
\end{figure}

We consider four data generating processes (DGPs) that differ in the distribution of the running variable as well as the underlying mean function $\mu$, and are illustrated in Figure \ref{dgp.paper2}. To generate values of the running variable $X_i$, we first draw independent values $Z_i$ from a particular Beta distribution, and then transform them using $X_i=2Z_i-1$ so that the support of the running variable is $[-1,1]$. The cutoff is set to be $c=0$. The response values $Y_i$ are generated from the values of the running variable $X_i$ as $Y_i=\mu(X_i)+\epsilon_i$, where $\epsilon_i \iid N(0, .1295^2)$. This is similar to the approaches taken in the simulation studies of CCT, IK, and SK. 

The first three DGPs are taken from the SK simulation. DGP1 is a modified version of a DGP from the simulation of AK. It has a flat Beta(1,1) distribution for $Z$ and a mean function consisting of quadratic splines with nodes not at the cutoff. DGP2 comes from the simulation of IK and is based on data from \textcite{lee2008randomized}. The variable $Z$ has a Beta(2,4) distribution, which means that approximately 19\% of the density is above the cutoff. The mean function consists of quintic polynomials on either side of the cutoff that differ only in their intercept. DGP3 is modeled after the Indiana school accountability data from SK that we also analyze in Section \ref{application}. It has a Beta(14,7) distribution for $Z$ with less than 6\% of its density below the cutoff, and a mean function consisting of separate cubic polynomials on either side of the cutoff. DGP4  uses a Beta(1,1) distribution for $Z$, but its underlying mean function is flat on either side of the cutoff, meeting the assumption of the local randomization method. The equations of the mean functions are 
\begin{align}
        \mu_{DGP1}(x)& =(x+1)^2-2s(x+0.2)+2s(x-0.2)-2s(x-0.4)+2s(x-0.7)-0.92 \nonumber \\
        & \quad +(0.1)\ind_{[x\geq 0]} \\
        \mu_{DGP2}(x)&=0.42+0.84x-3.0x^2+7.99x^3-9.01x^4+3.56x^5 + (0.1)\ind_{[x\geq 0]} \\
        \mu_{DGP3}(x)&=(0.05+1.5x+3.2x^2+2.7x^3)\ind_{[x<0]} + (0.15-0.15x+2.5x^2-1.5x^3)\ind_{[x\geq0]} \\
        \mu_{DGP4}(x)&=(0.1)\ind_{[x\geq0]}      
\end{align}
where $s(x)=(x)_+^2=$max$\{x,0\}^2$ is the square of the plus function. Each mean function has a discontinuity of 0.1.  Note that $\mu_{DGP1}$, $\mu_{DGP2}$, and $\mu_{DGP4}$ all satisfy Assumption \ref{sma}, so we might expect the PLE estimators to perform well in these simulations, whereas for $\mu_{DGP3}$ they may be inferior to the comparators that do not require smoothness at the cutoff.

Sample size and sparsity both contribute to the amount of information available for estimation at the cutoff. We consider different distributions of running variables because we are interested in the performance of RDD estimation methods when there are varying levels of sparsity near the cutoff. However, this makes it difficult to determine the samples sizes to use in such a simulation. For example, a sample of size 500 with a Beta(1,1) distribution yields approximately 250 values on either side of the cutoff, while that same sample size with a Beta(14,7) distribution yields only about 30 observations below the cutoff and 470 above the cutoff. The amount of information available near the cutoff is quite different for these scenarios. 

Thus, for setting up our simulation we employ the DISS metric $\bar{m}$ developed by SK, which is the expected number of observations within a rule of thumb bandwidth of the cutoff. Because this bandwidth is calculated based solely on the distribution of the running variable and is thus agnostic to the bandwidth algorithm or inferential method employed in the RDD estimation, we do not expect to systematically advantage one method over another by selecting sample size based on method-specific tailored algorithms.  SK demonstrate this advantage, as well as how the DISS metric promotes the ability to compare results across simulation settings with different running variable distributions. Table \ref{mn} gives the sample sizes for the values of $\bar{m}$ that we consider in our simulation study, identical to those in SK. We generate 50,000 simulated data sets for each combination of $\bar{m}$ and DGP. 

\begin{table}[t!]
\caption{Sample sizes for the different DGPs at each value of $\bar{m}$, which have been rounded to the nearest whole number.}
\begin{tabularx}{\textwidth}{lIIIII} 
\toprule
& $\bar{m}=10$ & $\bar{m}=21$ & $\bar{m}=27$ & $\bar{m}=44$ & $\bar{m}=57$ \\
\midrule
DGP1/DGP4 & 40 &  101 &  140  & 256  & 354  \\ 
DGP2 & 56 & 140  & 194  & 354  & 490  \\ 
DGP3 & 140  & 354  & 494  & 905  & 1254  \\ 
\bottomrule
\end{tabularx}
\label{mn}
\end{table}

\subsection{Simulation Study Results}
We evaluate the RDD estimation methods in terms of both point and interval estimation. One complication in doing so is the differing rates at which the estimation and interval calculation algorithms successfully produce finite numeric results with default settings. At the larger study sizes, there is near universal success. However, at the smaller study sizes, some of the simulated data sets are so sparse near the cutoff that some of the methods are unable to obtain finite estimates. For the sake of a fair comparison, the summaries presented in this section are calculated from the simulated data sets in which all methods obtained finite estimates for a particular DGP and study size. For $\bar{m}=10$, this represents between 90.38\% and 99.50\% of all simulated data sets. For the larger study sizes, this represents at least 99.99\% of all simulated data sets.

\subsubsection{Point Estimation}

\begin{figure}[t!]
    \centering
    \includegraphics[width=\textwidth]{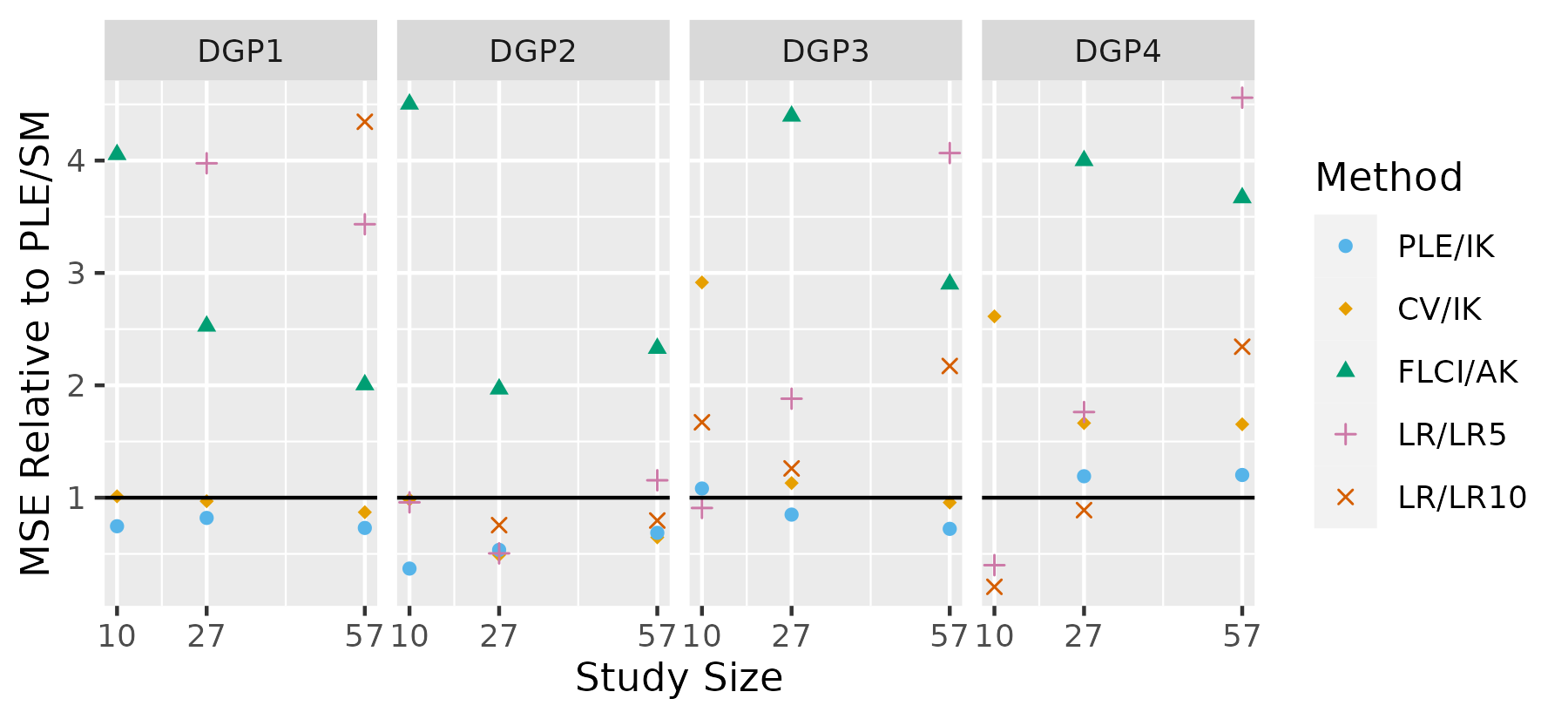}
    \caption{Mean squared error (MSE) for all methods at three study sizes, relative to PLE/SM.}
    \label{mse.paper2}
\end{figure}

Figure \ref{mse.paper2} depicts the mean squared error (MSE) values for the five comparator methods relative to our primary new method, PLE/SM, for the four DGPs at three of our study sizes ($\bar{m}$). See Figure \ref{mseb.paper2} in the Supplemental Appendix for MSE values at the other two study sizes. 

The relative performance of PLE/SM compared to the LR methods vary considerably based on the DGP and study size. The LR methods naturally work well in the flat DGP4 that satisfies the underlying local randomization assumption. However, for the other DGPs, the relative performance of the LR methods worsens as violations to the local randomization become more severe. This is primarily due to increasingly biased estimates without corresponding decreases in the empirical standard errors (EmpSEs), as seen in Figures \ref{bias.paper2}-\ref{empse.paper2} in the Supplemental Appendix. In DGP1 and in all but the smallest study size for DGP3, the PLE/SM method has smaller MSE values than all of the LR methods.

Compared to the continuity comparator methods, the PLE/SM method also works relatively well for the flat DGP4, with MSE values lower than all of these considered methods. All methods have very low bias for this DGP, and thus differences in MSE values are primarily due to EmpSE, in which the PLE methods perform relatively well. For the other DGPs, the results are more mixed. The MSE values of the PLS/SM methods are approximately half or less than those of FLCI/AK. But in many cases the CV/IK method produces MSE values lower than that of PLE/SM, particularly for the larger study sizes. 

The best performing method overall seems to be PLE/IK. It has the lowest MSE value for all study sizes in DGP1 and the majority of those in DGP3. In DGP2 it is either the best or nearly the best at all study sizes, and for DGP4 it is competitive in all but the smallest study size, where it has an inflated MSE value due to a single large outlier. 

\subsubsection{Interval Estimation}
\begin{figure}[t!]
    \centering
    \includegraphics[width=\textwidth]{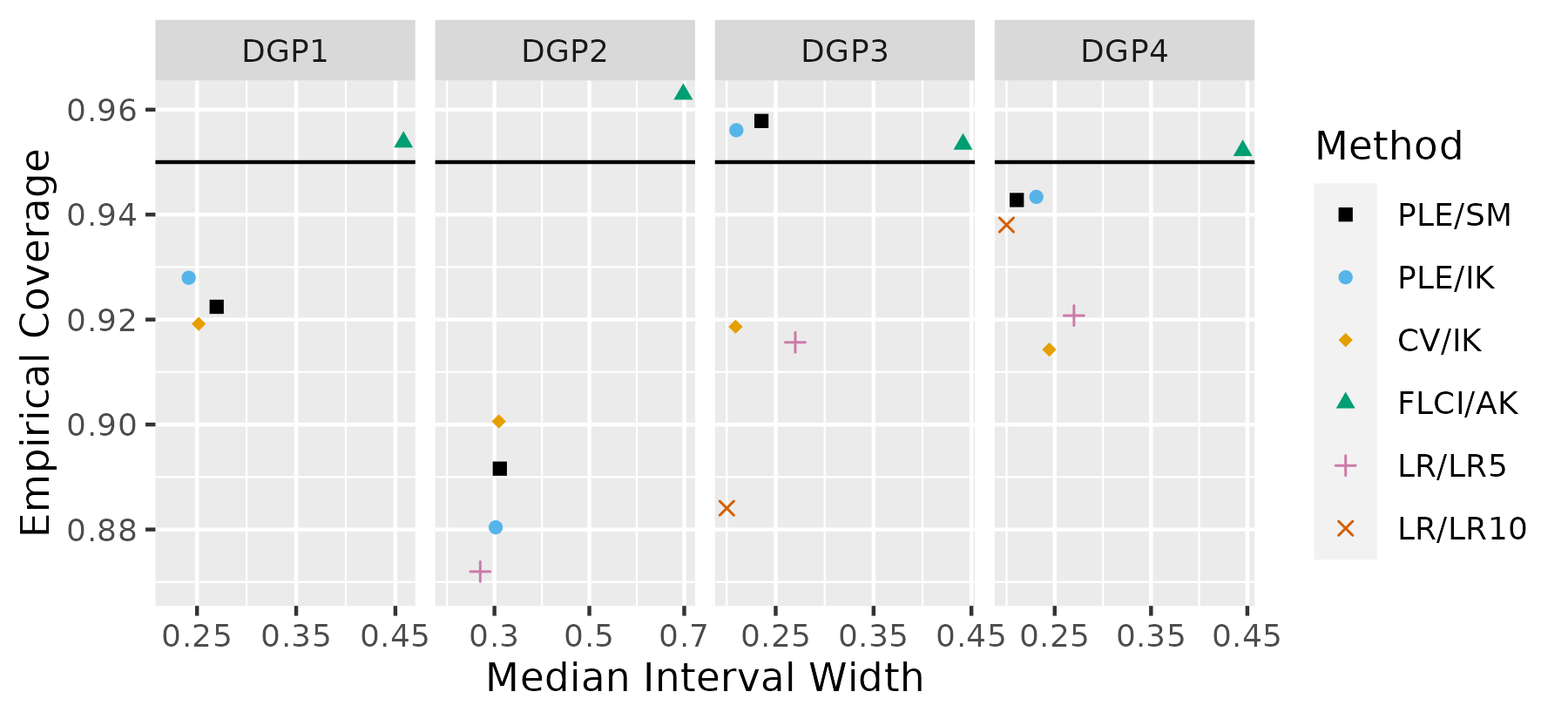}
    \caption{Median interval widths and empirical coverage for nominally 95\% confidence intervals for all methods at study size $m=27$. The graph omits the values of LR/LR10 for DGP1 (0.12) and DGP2 (0.61) as well as LR/LR5 for DGP1 (0.65). The Monte Carlo standard errors of the coverage estimates are approximately 0.001.}
    \label{coverage.paper2}
\end{figure}

We evaluate interval estimation in terms of the empirical coverage of the nominally 95\% confidence intervals produced by the different methods as well as the median width of the intervals, which are plotted in Figure \ref{coverage.paper2} for the middle study size of $\bar{m}=27$. Figures \ref{coverage.app.paper2}-\ref{coverage.app2.paper2} in the Supplemental Appendix give the same metrics at the other study sizes. Ideally we would like to see correct coverage produced by small intervals so the best methods will have a summary near the dark line and further left in each panel of the figures. 

Recall that the difference between the CV and FLCI methods presented here is that the latter inflates the critical value compared to the former as a way to address the issue of undercoverage in CV estimation. SK show this empirically in their simulation, with typically wider FLCI/AK intervals having better empirical coverage than narrower CV/IK intervals. As expected, the FLCI/AK method produces the widest intervals and often the highest coverage, while the CV/IK method produces narrower intervals with smaller but still somewhat reasonable coverage. The LR methods are competitive in DGP4 but less so in the other DGPs, especially DGP1 where the bias mentioned in the previous section leads to very poor coverage.

Both PLE methods perform quite well overall in terms of interval estimation. Their intervals are similar in width to the CV/IK intervals, but have better coverage in three of the four DGPs when $\bar{m}=27$. In DGP4, the PLE/IK method has better coverage and smaller median interval widths than CV/IK at all five study sizes, and the same is true for DGP1, except for $\bar{m}=10$. In DGP3, these two methods have virtually identical median widths, and yet PLE/IK has substantially higher coverage at each study size. At the larger three study sizes the PLE/IK intervals have essentially the same coverage as the FLCI/AK intervals. PLE/SM does not fair quite as well as PLE/IK yet is still competitive in several of the simulation settings. 

\subsubsection{Summary of Results}
Considering both point and interval estimation, the top performer in our simulation is the PLE/IK method. For DGP1 and DGP3, it outperforms both continuity and LR methods. Though the smoothness Assumption \ref{sma} does not hold for DGP3, the strengths of the PLE approach appear to outweigh the magnitude of the assumption violation in this particular example. For DGP2, PLE/IK outperforms the LR methods and is competitive with the continuity methods. Each of these represents a realistic scenario for an applied researcher, with DGP2 and DGP3 based on real data sets. DGP4 is a more contrived scenario that in many ways represents the best case for LR methods. PLE/IK is competitive there as well, although a more sophisticated window algorithm that could take advantage of more of the data may have given a much larger edge to the LR methods.

It is perhaps somewhat surprising that PLE inference tends to work better with the IK bandwidth algorithm than with SM, given the contradictory assumptions mentioned in Section \ref{skadevelop}. This certainly indicates that there are improvements to be sought in our SM bandwidth selection algorithm. In particular, the fact that SM relies on an asymptotic MSE formula is possibly problematic for small sample sizes. However, it also shows that the PLE method is somewhat robust to choice of bandwidth selection algorithm. 

\section{Data Application} \label{application}
\begin{figure}[t!]
    \centering
    \includegraphics[width=\textwidth]{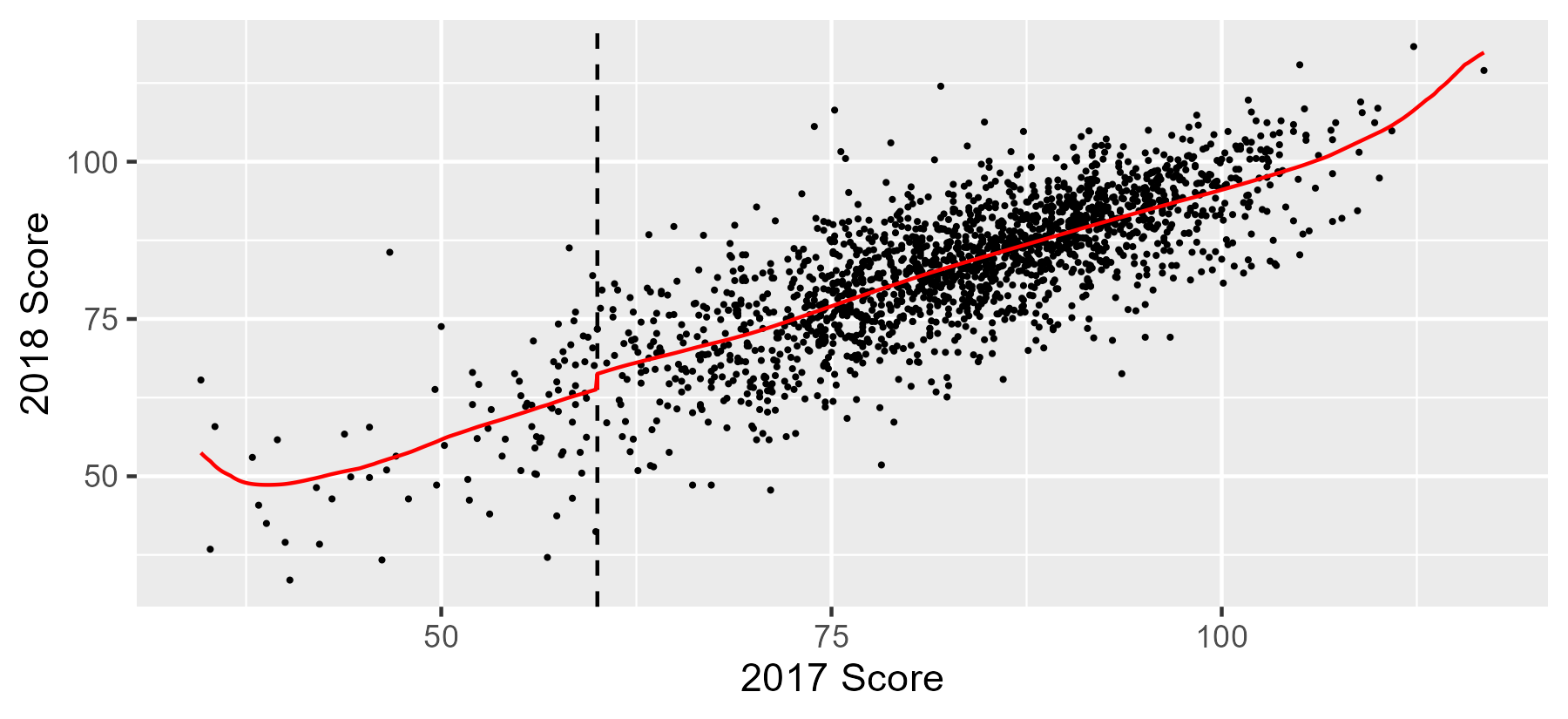}
    \caption{A scatterplot of the Indiana school accountability data. The dashed line represents the cutoff. The solid line represents the estimated mean function using the PLE/SM method, and the vertical part of that line at the cutoff represents the PLE/SM treatment effect estimate.}
    \label{scatter.paper2}
\end{figure}

We demonstrate the use of our partial linear estimator on school accountability scores from the state of Indiana, a data set first analyzed by SK. Beginning with the 2015-2016 school year, the Indiana Department of Education implemented a new accountability system in which each school was given a numeric score between 0 and 120. Schools that scored below a 60 were considered to be failing schools. The system outlined sanctions for schools that received failing grades multiple years in a row. The sanctions varied depending on the type of school and the number of consecutive failing years. We have data on these school accountability scores from the years 2017 and 2018, the first two years of full publicly available data (\textcite{Indianadata2017}, \textcite{Indianadata2018}). A scatterplot of the data is given in Figure \ref{scatter.paper2}.

This data set can be analyzed as a sharp RD design. The running variable is a school's score in the year 2017, and the response variable is the score in the year 2018. Since none of the sanctions take place after only one year of receiving a failing grade, the treatment is the threat of sanctions rather than the sanctions themselves. Presumably, the Indiana Department of Education would like the threat of sanctions to motivate failing schools to improve their scores in the following year. If that is true, than the scores should be higher slightly below the cutoff than they are slightly above. This would lead to a negative value of $\tau$ by the definition given earlier.

Out of the 1933 schools in the data set, 88 were given a failing score in 2017, representing less than five percent of the total. This is an example of a sparse data situation where the vast majority of the density of the running variable is far from the cutoff. There are $m=51$ values within a rule of thumb bandwidth of the cutoff and thus by the DISS metric falls between the $\bar{m}=44$ and $\bar{m}=57$ settings from our simulation, despite a larger overall sample size. 

Using our PLE/SM method, we estimate the average treatment effect to be 2.44, meaning that the threat of sanctions causes a decrease of 2.44 points in the following year's accountability score. This point estimate can be seen in Figure \ref{scatter.paper2}, along with the estimated mean function. This estimated treatment effect is indeed in the opposite direction of what the Department of Education would presumably like. A 90\% confidence interval for this effect is (-2.97,7.85), so we do not have statistically discernible evidence of an effect in either direction. If the Department of Education was hoping to increase school accountability scores by the threat of sanctions, it does not appear that their goal was met. However, this is a preliminary analysis without typical RD assumption checks, and thus limited in real world interpretability.

\begin{figure}[t!]
    \centering
    \includegraphics[width=\textwidth]{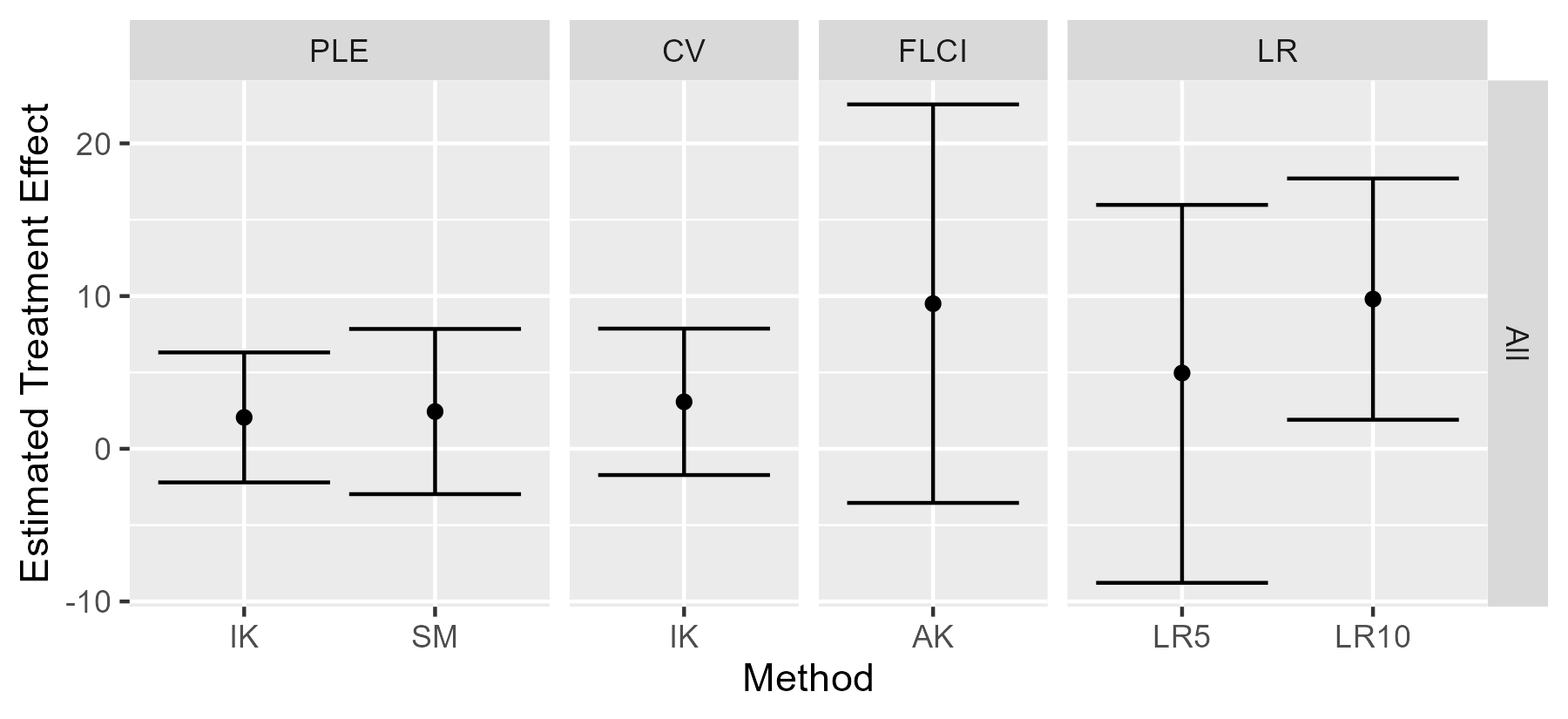}
    \caption{}
    \label{indiana.ci.paper2}
\end{figure}

Figure \ref{indiana.ci.paper2} gives point estimates and confidence intervals for this treatment effect from the other methods considered in our simulation study. The differences between methods here mirrors that of the simulation study. The PLE methods and the CV/IK method give relatively narrow intervals with similar centers. The FLCI/AK interval is more than twice as wide as the other continuity method intervals, fully containing them. It would appear that the local randomization assumption might not be reasonable in a very large window, leading to potential bias for the LR methods. This is particularly true when using the LR10 window, which results in a confidence interval completely above zero, but also a point estimate more than twice as high as that obtained by PLE and CV.  

\section{Conclusion} \label{conclusion}

Regression discontinuity designs are an important tool for applied researchers estimating causal effects of cutoff-based interventions in a variety of fields. Many of the methods that have been developed in recent years focus on asymptotic or large-sample properties, which likely led to researchers discounting the potential of the asymptotically inferior PLE approach. However, there are many applications with relatively small sample sizes or sparsity near the cutoff. There is a need for RD estimation methods that cater to these situations. Because PLE simultaneously incorporates points on both sides of the cutoff to estimate a single smooth mean function, PLE is a good candidate for small sample RD estimation. 

We introduced a new partial linear estimator that modifies Porter's original method to include local polynomial weights. We developed an MSE-optimal bandwidth algorithm that leverages an underlying smoothness assumption that is realistic in many RD settings. We paired our estimator with a jackknife-based variance estimation technique for inference. We showed that the resulting PLE/SM method is always competitive with and often outperforms certain popular methods in a set of small sample data scenarios. In our simulations the top performing method was our PLE approach paired with the IK bandwidth selection algorithm.

The methods we developed have a strong potential to help researchers working with small sample RD designs. We further hope that our work sparks renewed interest in small study RD estimation methods. In the future we plan to refine our bandwidth algorithm to improve its performance, and to extend our methodology to fuzzy RD designs and other, more complicated, design structures.

\newpage

\printbibliography

\newpage

\section*{Supplemental Appendix}
\renewcommand{\thefigure}{A\arabic{figure}}
\renewcommand{\thesection}{A\arabic{section}}
\setcounter{section}{0}  
\setcounter{figure}{0}  
\section{Additional Figures}
\newpage

\begin{figure}[t!]
    \centering
    \includegraphics[width=\textwidth]{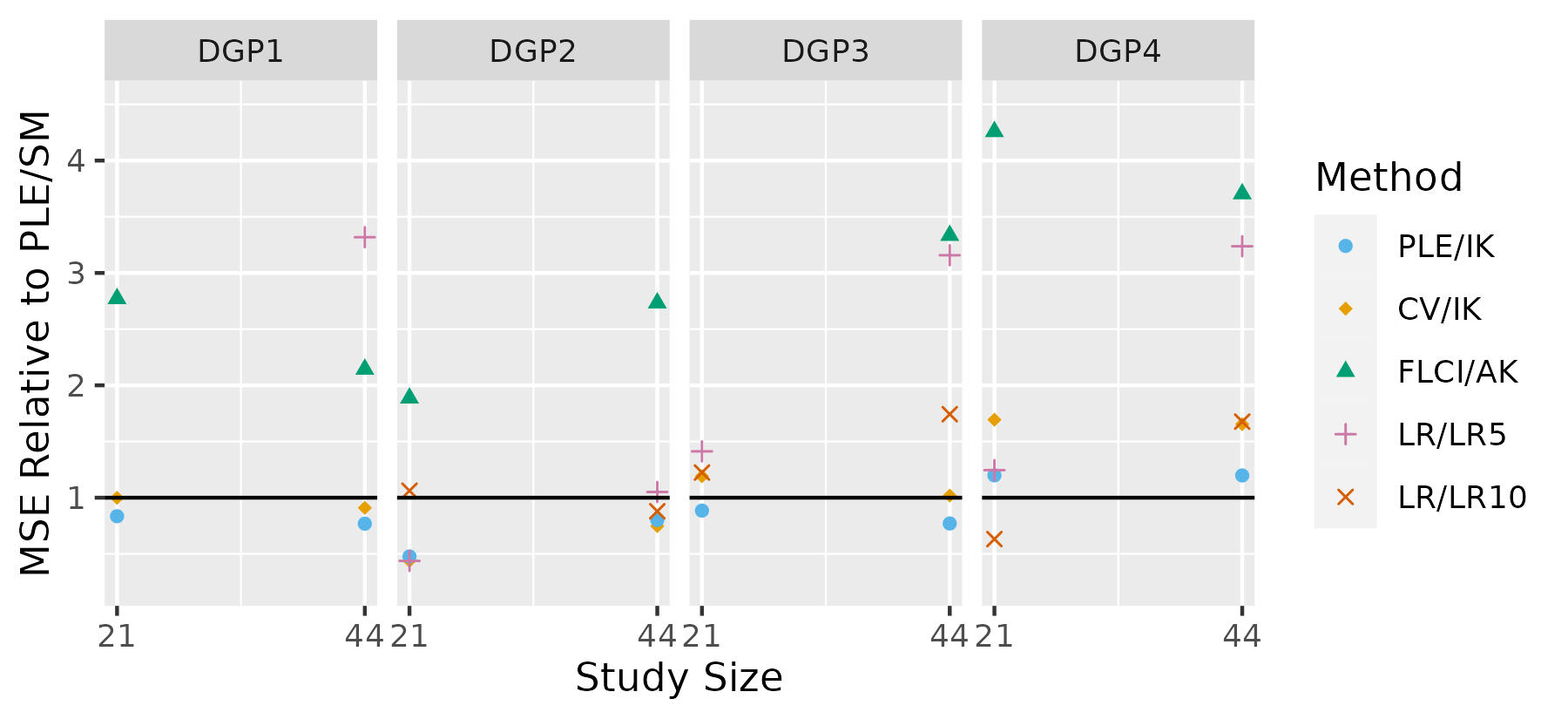}
    \caption{Mean squared error (MSE) for all methods at two study sizes, relative to PLE/SM.}
    \label{mseb.paper2}
\end{figure}

\begin{figure}[t!]
    \centering
    \includegraphics[width=\textwidth]{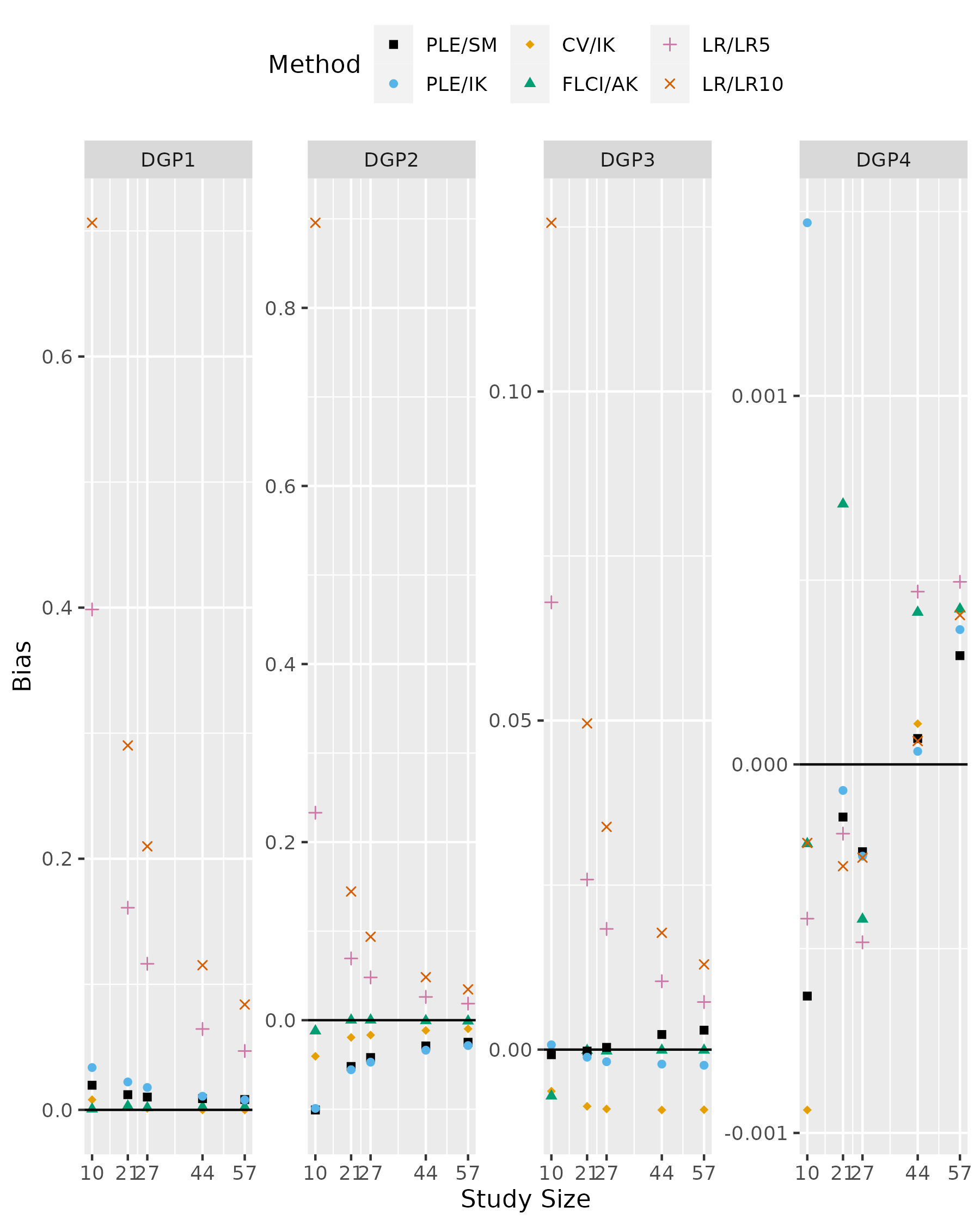}
    \caption{Bias for all methods at all study sizes. The Monte Carlo standard errors of the estimates shown are at most 0.001 except for $\bar{m}=10$ values of FLCI/AK (0.02 for DGP3, 0.003 for DGP2), PLE/IK (0.003 for DGP4), and LR/LR10 (0.003 for DGP2).}
    \label{bias.paper2}
\end{figure}

\begin{figure}[t!]
    \centering
    \includegraphics[width=\textwidth]{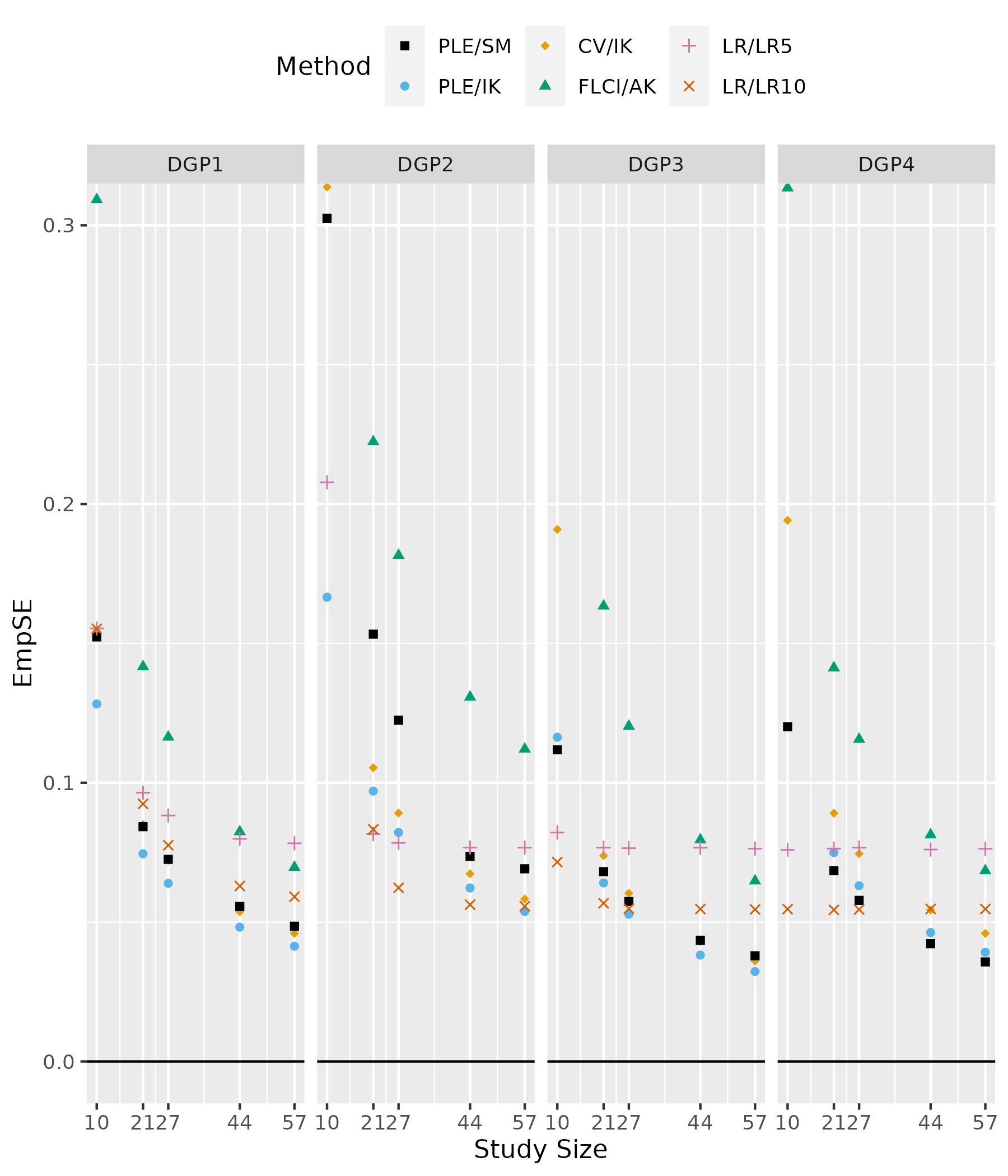}
    \caption{Empirical standard errors (EmpSE) for all methods at all study sizes. The graph omits values when $\bar{m}=10$ for FLCI/AK (4.2 for DGP3, 0.7 for DGP2), PLE/IK (0.8 for DGP4), and LR/LR10 (0.7 for DGP2). The Monte Carlo standard errors of the estimates shown are at most 0.001.}
    \label{empse.paper2}
\end{figure}

\begin{figure}[t!]
    \centering
    \includegraphics[width=\textwidth]{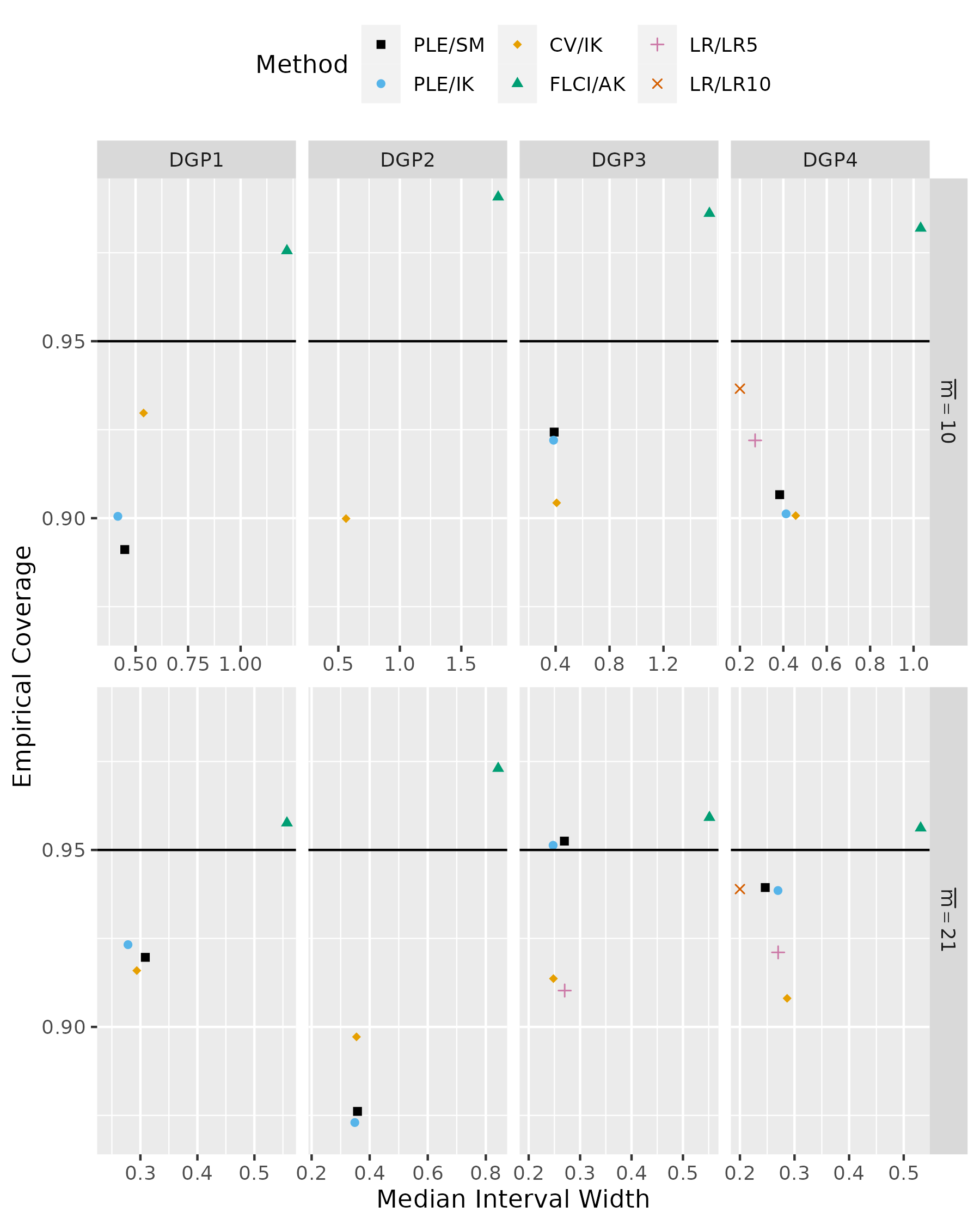}
    \caption{Median interval widths and empirical coverage for nominally 95\% confidence intervals for two study sizes. Omitted values have coverage lower than 0.87. The Monte Carlo standard errors of the coverage estimates shown are at most 0.001.}
    \label{coverage.app.paper2}
\end{figure}

\begin{figure}[t!]
    \centering
    \includegraphics[width=\textwidth]{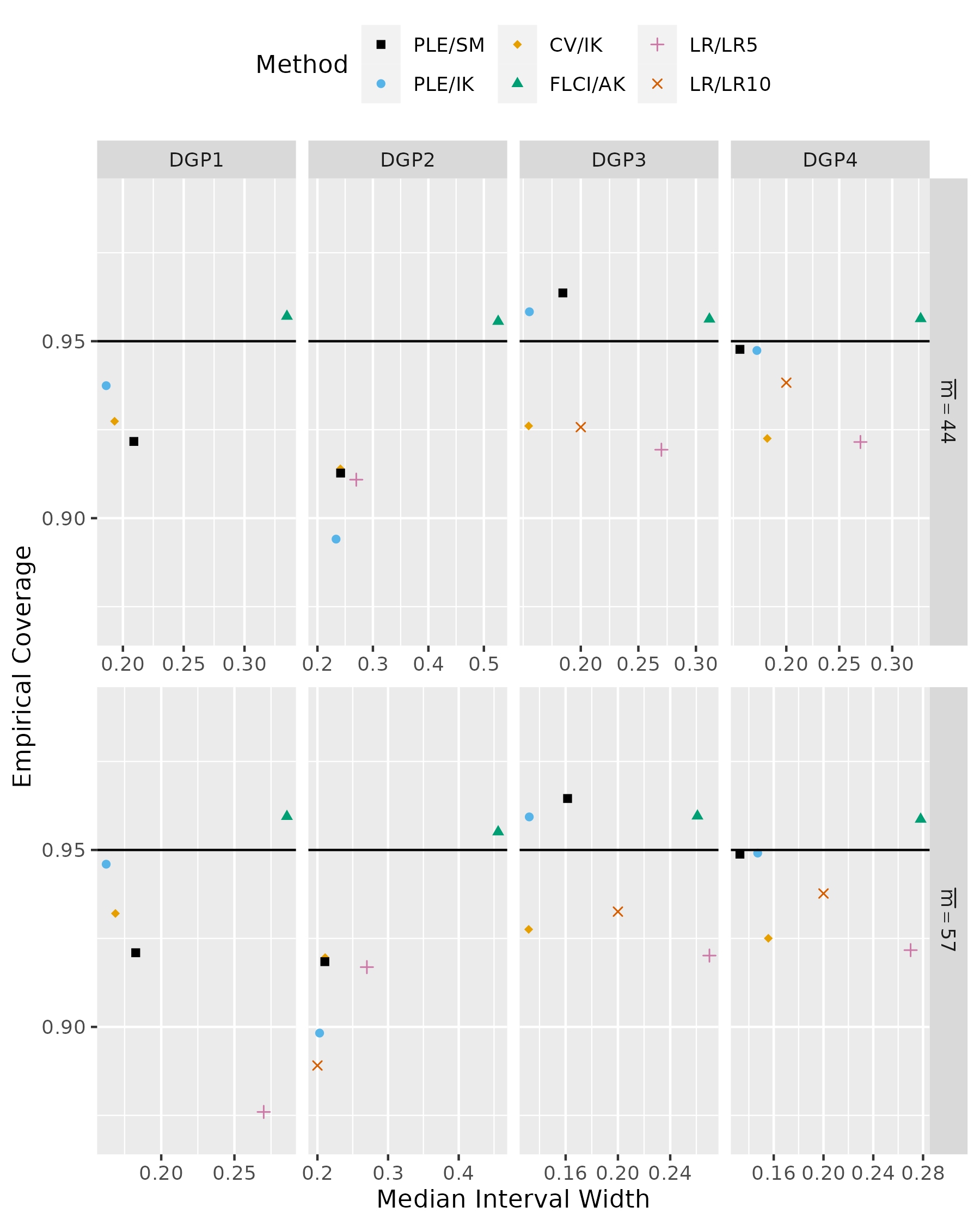}
    \caption{Median interval widths and empirical coverage for nominally 95\% confidence intervals for two study sizes. Omitted values have coverage lower than 0.87. The Monte Carlo standard errors of the coverage estimates shown are at most 0.001.}
    \label{coverage.app2.paper2}
\end{figure}

\clearpage

\section{Extended Simulation Results}
In this section we present results from an extended simulation comparing RD estimation methods. Here we consider the PLE method but with $p=0$ corresponding to local constant weights rather than the local linear weights in the main paper. We refer to this as PLE0. The SM bandwidth is calculated from the formula with $p=0$, and perhaps for the small study sizes we are interested in a local constant estimator may have some advantages over a local linear estimator. We pair PLE0 with both the SM and IK bandwidths.

In addition, we consider the same local randomization method as before, but in this case we set a minimum of up to 20 in the window, referred to as LR20, instead of minimums of five and ten as in the main paper. We expect that for the larger study sizes considered, and for the flat DGP4, including more observations in the window will lead to better results. 

For reference, we include the PLE/SM method in these figures as well.
Note that these results are based on the simulated data sets in which all of these additional methods are able to provide an interval estimate, and thus there are minute differences in the PLE/SM values here compared to the main paper for the smaller study sizes. 

\begin{figure}[t!]
    \centering
    \includegraphics[width=\textwidth]{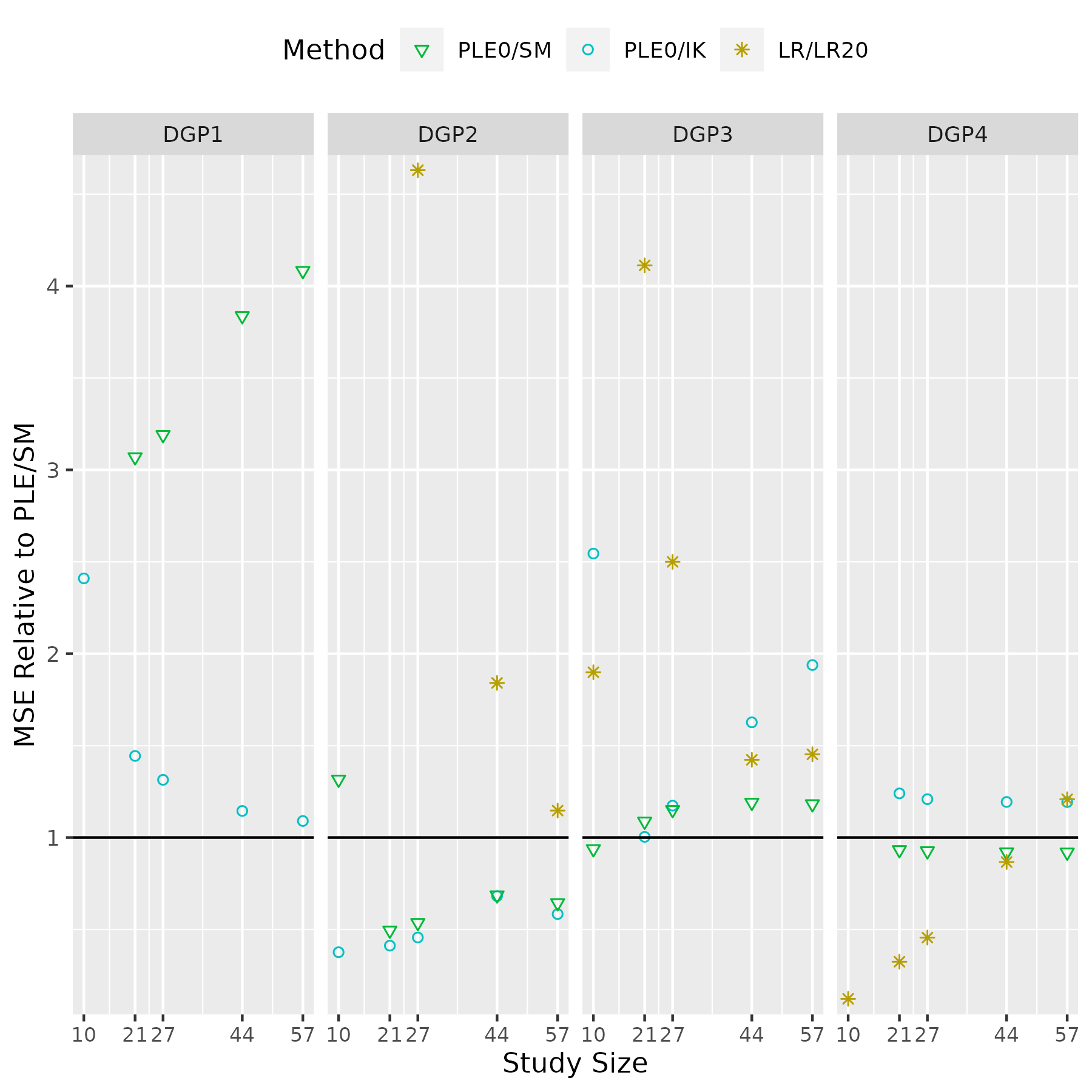}
    \caption{Mean squared error (MSE) for additional methods at all study sizes, relative to PLE/SM.}
    \label{mse.paper2app}
\end{figure}

In Figure \ref{mse.paper2app} we see that in DGP1 and DGP3, the PLE0 methods typically perform slightly worse than their PLE counterparts. For DGP2 the PLE0 methods perform relatively better, although the PLE0/IK methods are quite similar to the PLE/IK methods presented in the main paper. Again we see that in DGP4 the SM bandwidth leads to better results than the IK bandwidth, but that for smaller study sizes LR/LR20 performs the best. 

\begin{figure}[t!]
    \centering
    \includegraphics[width=\textwidth]{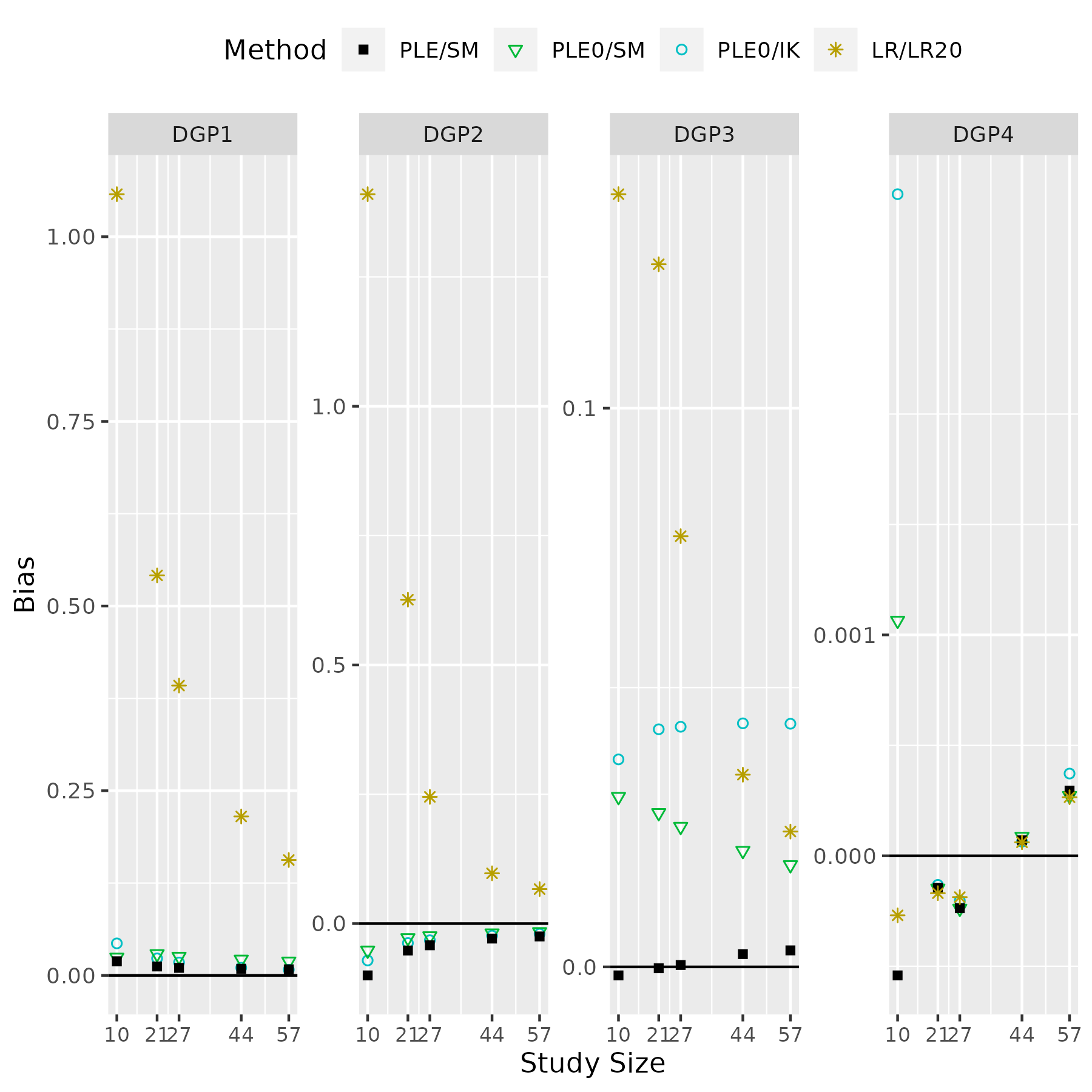}
    \caption{Bias for additional methods at all study sizes. The Monte Carlo standard errors of the estimates shown are at most 0.001 except for $\bar{m}=10$ values of PLE0/IK (0.01 for DGP4), PLE0/SM (0.006 for DGP1, 0.002 for DGP2 and DGP4), and LR/LR20 (0.003 for DGP2), as well as $\bar{m}=21$ value of LR/LR20 (0.002 for DGP2).}
    \label{bias.paper2app}
\end{figure}

\begin{figure}[t!]
    \centering
    \includegraphics[width=\textwidth]{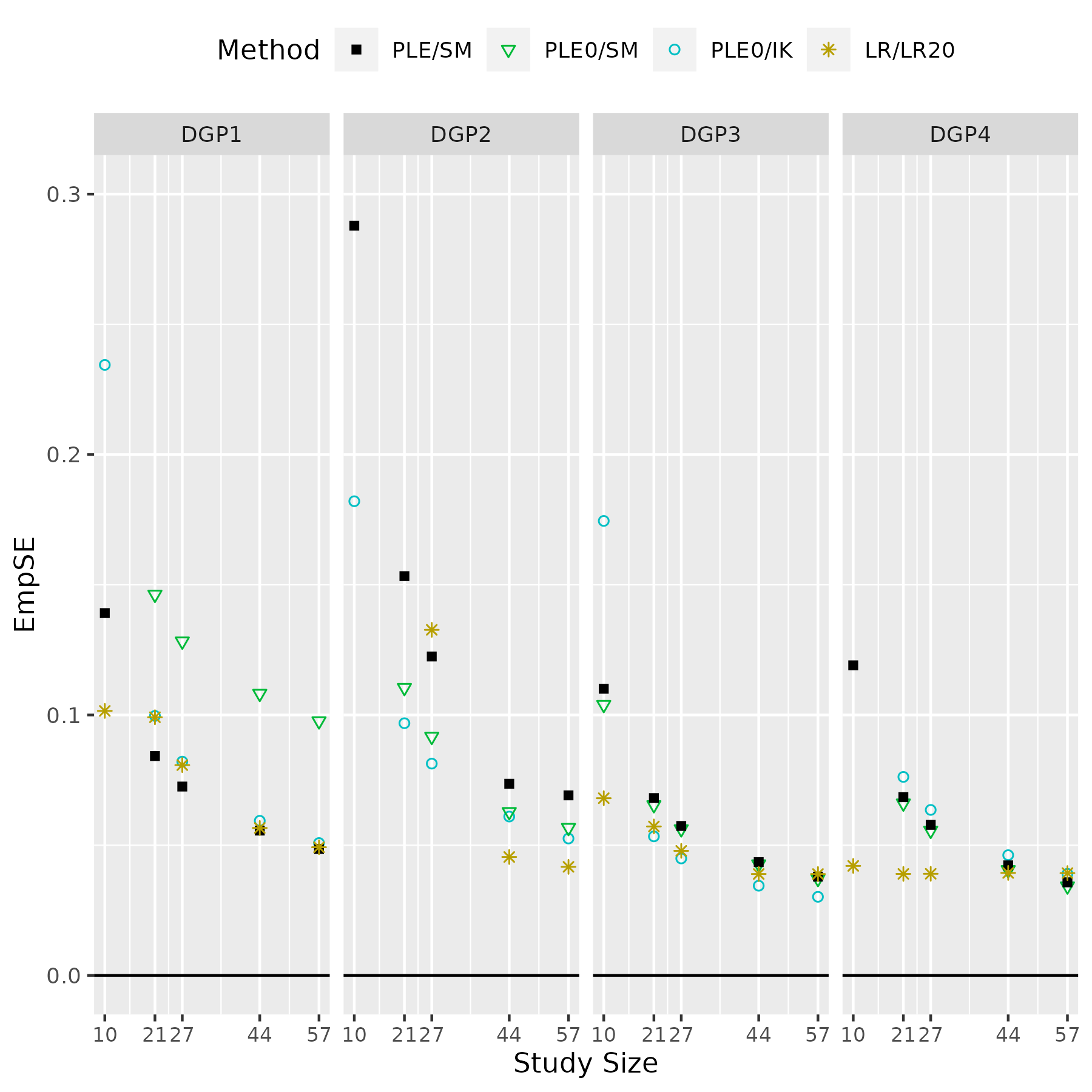}
    \caption{Empirical standard errors (EmpSE) for additional methods at all study sizes. The graph omits values when $\bar{m}=10$ for PLE0/IK (2.2 for DGP4), PLE0/SM (1.3 for DGP1, 0.4 for DGP2 and DGP4) and LR/LR20 (0.7 for DGP2) and when $\bar{m}=21$ for LR/LR20 (0.5 for DGP2). The Monte Carlo standard errors of the estimates shown are at most 0.001.}
    \label{empse.paper2app}
\end{figure}

Figures \ref{bias.paper2app}-\ref{empse.paper2app} give the bias and EmpSE values for these additional methods, and again a very similar pattern emerges as in the main paper. The LR/LR20 method has extreme bias, other than in DGP4, that only slightly improves at our largest study sizes. The poor performance of PLE0/SM is due to larger bias in DGP3, but larger EmpSE in DGP1. 

\begin{figure}[t!]
    \centering
    \includegraphics[width=\textwidth]{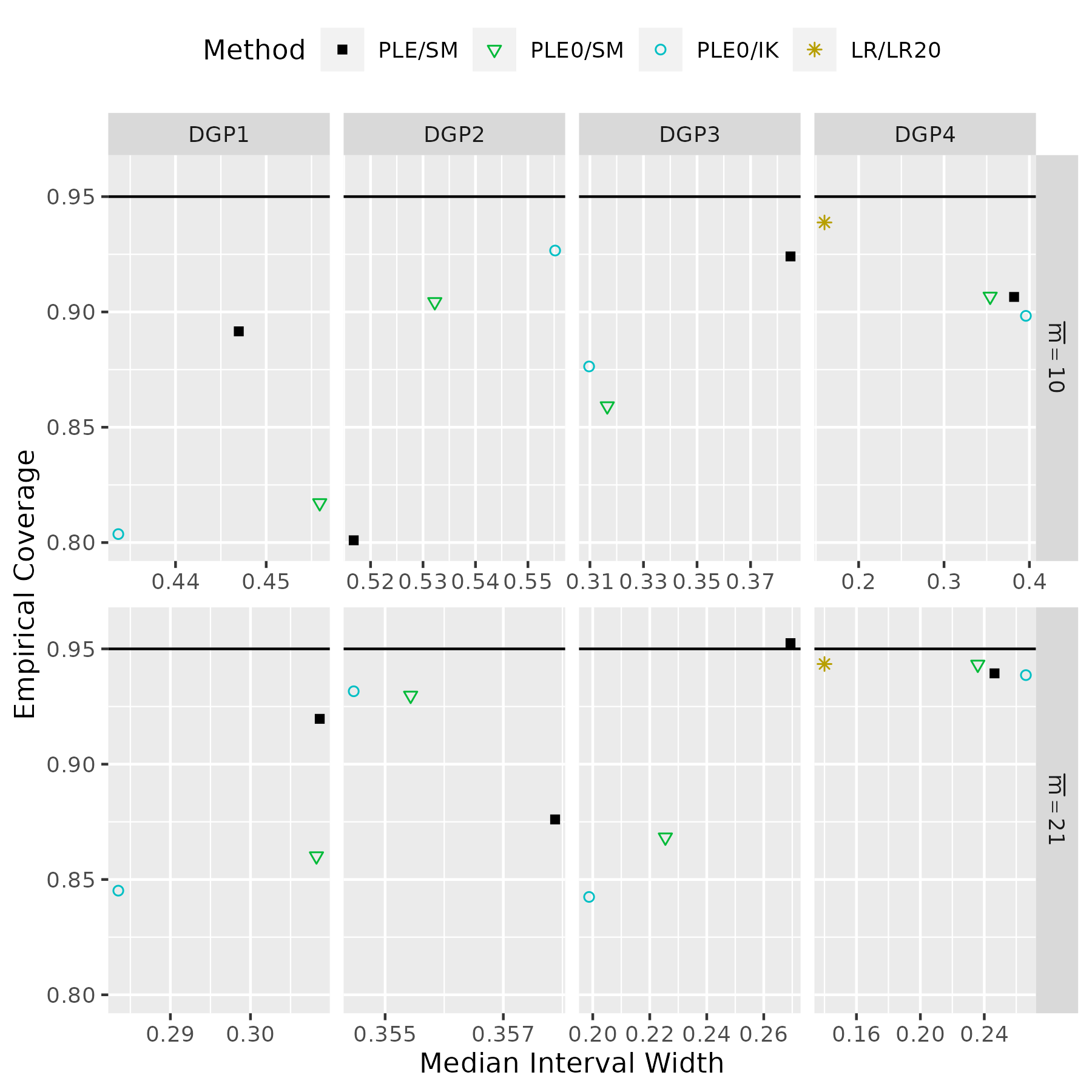}
    \caption{Median interval widths and empirical coverage for nominally 95\% confidence intervals for two study sizes. The graph omits LR/LR20 values when $\bar{m}=10$ for DGP1 (width of 0.36, coverage of 0), DGP2 (1.1,0.0003), and DGP3 (0.20, 0.32) and when $\bar{m}=21$ for DGP1 (0.22,0), DGP2 (0.27,0.001), and DGP3 (0.14, 0.17). The Monte Carlo standard errors of the coverage estimates shown are at most 0.002.}
    \label{coverage.paper2app}
\end{figure}

\begin{figure}[t!]
    \centering
    \includegraphics[width=\textwidth]{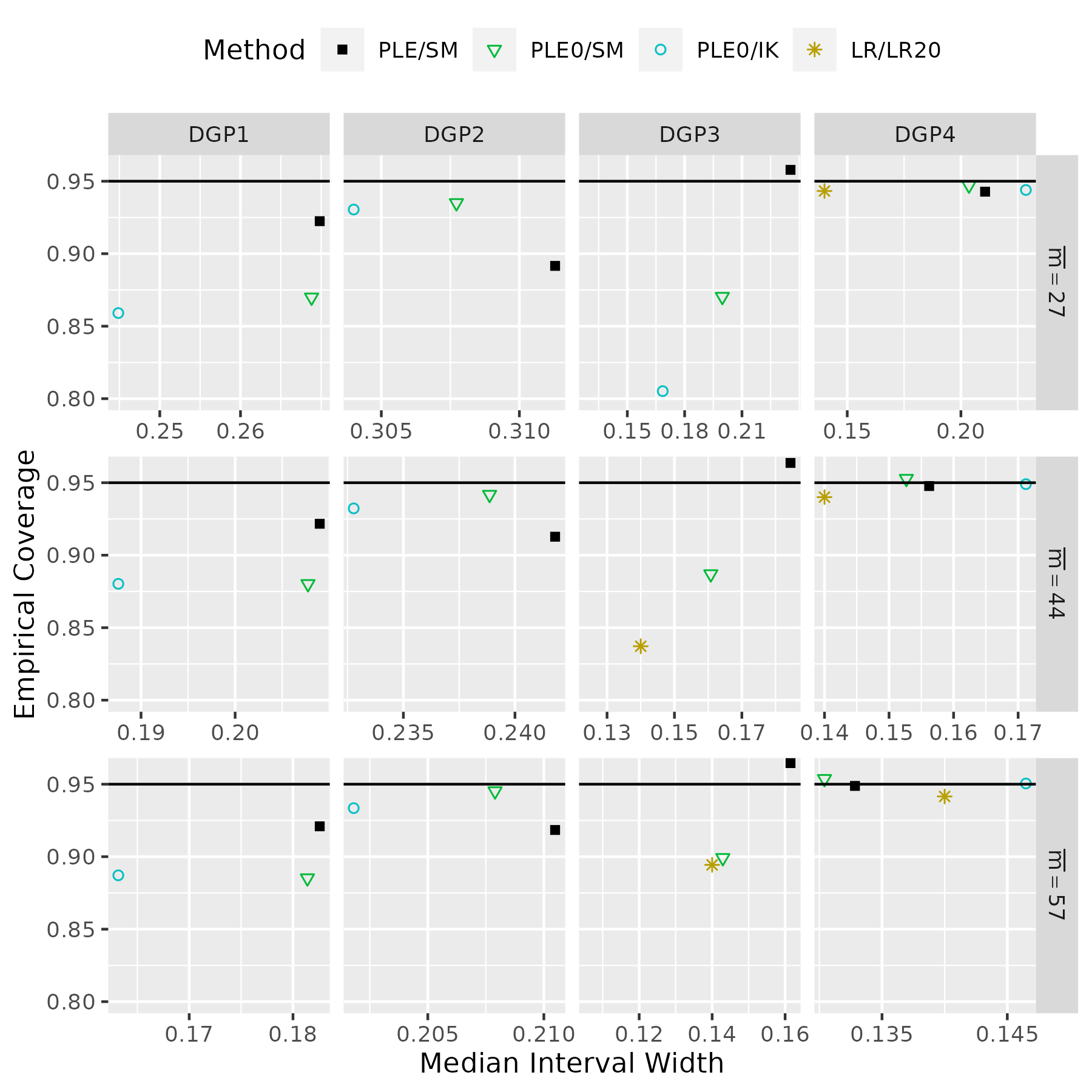}
    \caption{Median interval widths and empirical coverage for nominally 95\% confidence intervals for three study sizes. The graph omits LR/LR20 values when $\bar{m}=27$ for DGP1 (width of 0.2, coverage of 0), DGP2 (0.18,0.02), and DGP3 (0.13, 0.47), when $\bar{m}=44$ for DGP1 (0.16,0.01) and DGP2 (0.14,0.34), and when $\bar{m}=57$ for DGP1 (0.15,0.07) and DGP2 (0.14,0.6). The Monte Carlo standard errors of the coverage estimates shown are at most 0.002.}
    \label{coverage.paper2app2}
\end{figure}

The three continuity methods considered here lead to very similar interval widths, as seen in Figures \ref{coverage.paper2app}-\ref{coverage.paper2app2}. Again the PLE0 methods struggle in DGP1 and DGP3, here in terms of coverage, but do slightly better in DGP3. The LR/LR20 intervals lead to severe undercoverage for small studies with non-flat mean functions.

Overall, these results indicate that there may be some situations when using PLE0 instead of PLE can lead to improved estimation. However, it more often led to poor estimates in the realistic DGPs studied here, while PLE1 remained competitive in these situations as well. Due to this we feel comfortable recommending PLE use in general, although future simulations with other settings may shed more light on the issue.

\clearpage

\section{Variance Estimation} \label{appvar}

This section contains several alternatives to the jackknife variance estimation technique described in \ref{plevariance}, and results from a simulation study comparing these techniques.

\subsection{Alternative Techniques}

The jackknife is a resampling procedure that typically removes one observation at a time from a data set, calculates an estimate based on the remaining $n-1$ observations, and then utilizes those $n$ point estimates to estimate the variance of the overall estimate. Recall from \ref{plevariance} that the our chosen variance estimation technique $\hat{V}_{PLE}$ is based on a formula from \textcite{wu1986jackknife}. We also consider a similar method based on the formula from \textcite{hinkley1977jackknifing},

\begin{equation} \label {hinkleyresid r}
    \hat{V}_{H}(\hat{\tau})=(\hat{\D}'\hat{\D})^{-1}\sum_{i=1}^n\frac{r_i^2}{1-1/n}\hat{d}_i^2(\hat{\D}'\hat{\D})^{-1}
\end{equation}

where $\hat{d}_i=d_i-\hat{E}(d|x_i)$, $\hat{\D}=(\hat{d}_1,...\hat{d}_n)'$, and $r_i=y_i-\hat{E}(y|x_i)-\hat{d}_i\hat{\tau}$.

Note that the bandwidth calculation so crucial for RD analysis factors into the estimation of $E(Y|X)$ and $E(D|X)$ and thus the residuals. However, once the residuals have been calculated the bandwidth plays no further role in the estimation. Thus removing one pair of residuals each time to recalculate the treatment effect estimate does not necessitate recalculating the bandwidth or the nonparametric estimates. This leads to a potential concern, that by applying the jackknife at the second step of the partially linear estimation, after the nonparametric regression has taken place, we may be underestimating the variance of our estimate. Therefore as an alternative to the above jackknife approach that deletes one pair of residuals each time, we consider a method that deletes an original observation each time. In this case, since the subsequent step involves local polynomial regression, methods based on parametric regression such as $\hat{V}_{PLE}$ and  $\hat{V}_{H}$ cannot be directly applied.  However, we can still get an estimate $\hat{\tau}$ based on the full set of data and estimates $\hat{\tau}_{(i^*)}$ based on removing the $i$th observation. Using those estimates in the Hinkley and Wu formulas yields variance estimates

\begin{equation} \label {hinkleyorig}
    \hat{V}_{H,O}(\hat{\tau})=\frac{1}{n(n-1)}\sum_{i=1}^n\left(n(1-w_i)(\hat{\tau}-\hat{\tau}_{(i^*)})\right)^2
\end{equation}
and 
\begin{equation} \label {wuorig}
    \hat{V}_{W,O}(\hat{\tau})=\sum_{i=1}^n(1-w_i)(\hat{\tau}_{(i^*)}-\hat{\tau})^2,
\end{equation}
where $w_i=\hat{d}_i'(\hat{\D}'\hat{\D})^{-1}\hat{d}_i$ and the $O$ in the subscript indicates that we are deleting an original observation. While these methods lack the theoretical justification and proof of asymptotic properties enjoyed by $\hat{V}_{PLE}$ and $\hat{V}_{H}$, they represent a natural extension of the jackknife for a parametric regression into the nonparametric framework.

In addition to the jackknife methods, we consider the following plug-in variance estimator from Porter based on Theorem \ref{asymp dist}:
\begin{equation}
    \hat{V}_P(\hat{\tau})=\frac{C_{P1}(\hat{\sigma}^{2}_+(c)+(\hat{\sigma}^{2}_-(c))}{4\hat{f}(c)nh}.
\end{equation}

Finally, recall from Equation \ref{tauhatple} that we can write the PLE estimate in a linear smoother form as $\hat{\tau}^{PLE}=\mathbf{(G'G)}^{-1}\mathbf{G'(I-L')Y}$, which has a constant matrix multiplied by the random vector $Y$. Thus we can apply the formula for the variance of a linear transformation to get
\begin{equation}
    Var(\hat{\tau}^{PLE})=\mathbf{(G'G)}^{-1}\mathbf{G'(I-L')}\mathbf{\Sigma} [\mathbf{(G'G)}^{-1}\mathbf{G'(I-L')}]',
\end{equation}
where $\mathbf{\Sigma}=Var(\mathbf{Y})=diag(\sigma_1^2,..., \sigma_n^2)$. The obvious challenge here is that we do not know $\mathbf{\Sigma}$ and therefore  must estimate it from the data. For this we use a modification of the nearest neighbors approach from CCT. The modification is necessary because the nearest neighbors to a particular observation may be on the other side of the cutoff. Therefore we subtract our estimate of $\tau$ from the $y$ value of all points above the cutoff, yielding an estimate of 
\begin{equation}
    \hat{\sigma}_{i,NN}^2=\frac{J}{J+1}\left((Y_i-\hat{\tau}^{PLE}\ind_{[x_i\geq c]})-\frac{1}{J}\sum_{j=1}^J\left[Y_{j(i)}-\hat{\tau}^{PLE}\ind_{[x_{j(i)]}\geq c}\right]\right)^2,
\end{equation}
where $J$ is the number of nearest neighbors and $Y_{j(i)}$, $x_{j(i)}$ are the $Y$ and $x$ values, respectively, of the observation whose $x$ value is the $j$th nearest to $x_i$. This leads to a direct plug-in (DPI) variance estimate
\begin{equation}
    \hat{V}_{DPI}(\hat{\tau})=\mathbf{(G'G)}^{-1}\mathbf{G'(I-L')} diag(\hat{\sigma}_{1,NN}^2,..., \hat{\sigma}_{n,NN}^2) [\mathbf{(G'G)}^{-1}\mathbf{G'(I-L')}]'.
\end{equation}

\subsection{Simulation Settings} \label{variance sim}
We use a Monte Carlo simulation study to compare the variance estimators mentioned above. Based on these results, we recommend using the $\hat{V}_{PLE}$ formula which incorporates a jackknife technique that leaves out one of the residuals each time.

The simulation here is similar in settings to that in the main paper, with a slightly more limited scope. We consider only the first three DGPs and the PLE method with the SM and IK bandwidths. We implement a sparsity adjustment for these bandwidths by calculating the minimum bandwidth that will produce a finite PLE estimate and increasing the bandwidth if it is not at least that large. Because of this adjustment, the methods are successful for all generated data sets. We use the same study sizes as in the main paper, and generate 10000 data sets for all but DGP3 for $\bar{m}=57$, which has 3700 data sets. 

We are interested here in evaluating the performance of the variance estimation formulas, not the PLE estimation method as a whole. Thus at each study size and DGP, we estimate the average model standard error for each method, 
 \begin{equation}
     ModSE=\sqrt{\frac{1}{n}\sum_{i=1}^n V(\hat{\tau}_i)}.
 \end{equation}
 We also estimate the empirical standard error,
 \begin{equation}
     EmpSE=\sqrt{\frac{1}{n}\sum_{i=1}^n (\hat{\tau}_i-\bar{\hat{\tau}})^2)}.
 \end{equation}
 which serves as an estimate for the true standard error of the RD treatment effect estimate, and does not depend on the variance estimation technique. We compare these two with an estimate of the relative percent error in the model standard error, 
 \begin{equation}
     RelE=100\left(\frac{ModSE}{EmpSE}-1\right).
 \end{equation}
Positive values of this relative error indicate that an estimation technique is overestimating the true variance, while negative values indicate underestimation. We calculate Monte Carlo standard errors (MCSE) of these three estimates using the formulas from \textcite{morris2019using}. 

\subsection{Simulation Results} \label{variance sim results}

\begin{figure}[t!]
    \centering
    \includegraphics[width=\textwidth]{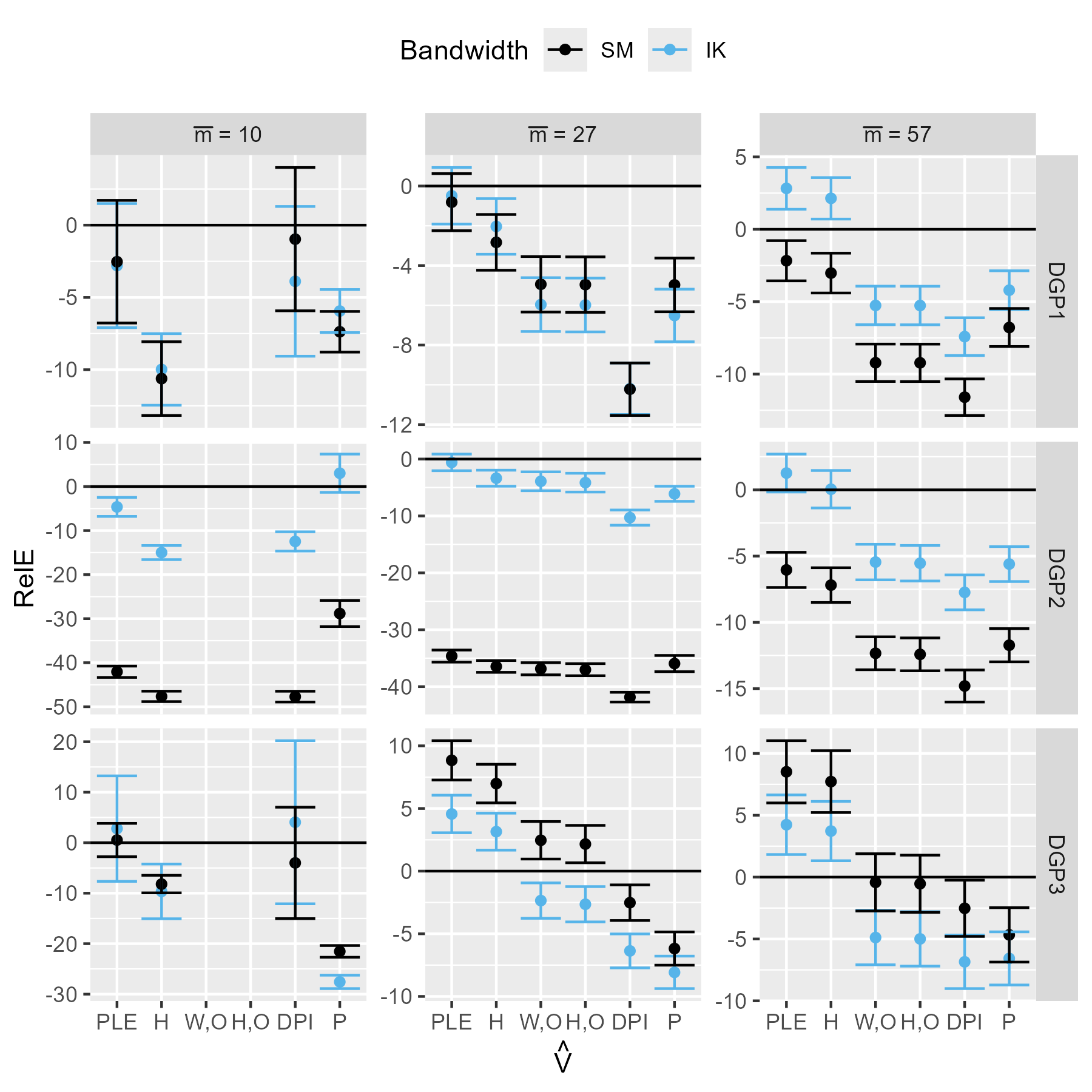}
    \caption{RelE values for all variance estimators using PLE at certain study sizes. The graph omits values of $\hat{V}_{W,O}$ and $\hat{V}_{H,O}$ for $\bar{m}=10$ (all above 150). Error bars represent 1.96 MCSE above and below the point estimate.}
    \label{relerror.paper2}
\end{figure}

Figure \ref{relerror.paper2} gives RelE values for the different techniques at study sizes of 10, 27, and 57. The variance estimation techniques with the overall best performance in these settings in terms of RelE are those that delete a residual in each jackknife procedure, $\hat{V}_{PLE}$ and $\hat{V}_{H}$. These techniques typically outperform the asymptotic $\hat{V}_P$ for small studies and often do so for larger studies as well. The $\hat{V}_{DPI}$ technique is also competitive in certain situations, but in others it performs quite poorly. One concern with the DPI approach is that the estimated variance-covariance matrix could be correlated with the treatment effect estimate, which may be contributing to the inconsistent performance we see here. The $\hat{V}_{PLE}$ and $\hat{V}_{H}$ techniques, on the other hand, are almost universally competitive.

These jackknife methods that delete a residual each time also avoid the very poor performances exhibited by $\hat{V}_{H,O}$ and $\hat{V}_{W,O}$ for small study sizes. These latter methods, which delete an original observation in the jackknife procedure, severely overestimate the true variance on average in our small study settings, and do not show a substantial improvement over $\hat{V}_{PLE}$ and $\hat{V}_{H}$ even when they give reasonable estimates for larger study sizes. This poor performance is not merely the case of a handful of iterations producing extreme standard error estimates. Rather, a substantial percentage of iterations produce high values for these methods. For small study sizes there is seemingly too much variation involved in deleting an original observation in the jackknife procedure. Our original concern that $\hat{V}_{PLE}$ and $\hat{V}_{H}$ would not be able to capture the true variability, which led to the proposal of $\hat{V}_{H,O}$ and $\hat{V}_{W,O}$, does not appear to be valid. Conversely, it seems that by deleting a residual in the jackknife procedure we are able to maintain reasonable standard error estimates. Considering their poor performance and the lack of theoretical justification, we do not recommend the use of $\hat{V}_{H,O}$ and $\hat{V}_{W,O}$ in practice.

The performances of the top two techniques, $\hat{V}_{PLE}$ and $\hat{V}_{H}$, are fairly similar in most settings. We propose the use of $\hat{V}_{PLE}$ for two primary reasons. One is the justification provided by \textcite{wu1986jackknife} in comparing their method to that of $\hat{V}_{H}$. The second is that, at least in the settings in this simulation study, $\hat{V}_{PLE}$ tends to be slightly larger than $\hat{V}_{H}$. Considering that RD confidence intervals other than those using FLCI tend towards undercoverage, having a slightly larger variance may lead to better results in practice.

\end{document}